\documentclass[conference]{IEEEtran}
\ifCLASSINFOpdf
\else
\fi
\usepackage{cite}
\usepackage{graphicx}
\usepackage{listings}
\usepackage{listings-rust}
\usepackage[table]{xcolor}
\usepackage{float}
\usepackage{booktabs}
\usepackage{array} 
\usepackage{makecell}

\definecolor{pastelred}{RGB}{255, 179, 186}
\definecolor{pastelgreen}{RGB}{186, 255, 201}

\usepackage{subfig}

\begin{document}


%
\title{NPB-Rust: NAS Parallel Benchmarks in Rust}

 \author{
 \IEEEauthorblockN{Eduardo M. Martins, Leonardo Gibrowski Faé, Renato B. Hoffmann, Lucas S. Bianchessi, Dalvan Griebler}
 \IEEEauthorblockA{School of Technology, Pontifical Catholic University of Rio Grande do Sul (PUCRS)\\
 Parallel Application Modeling Group (GMAP), Porto Alegre – RS – Brazil\\
 Email: (e.martins01, leonardo.fae, renato.hoffmann, l.bianchessi)@edu.pucrs.br, dalvan.griebler@pucrs.br}
 }


%


\maketitle


\begin{abstract}


Parallel programming often requires developers to handle complex computational tasks that can yield many errors in its development cycle. Rust is a performant low-level language that promises memory safety guarantees with its compiler, making it an attractive option for HPC application developers. We identified that the Rust ecosystem could benefit from more comprehensive scientific benchmark suites for standardizing comparisons and research. The NAS Parallel Benchmarks (NPB) is a standardized suite for evaluating various hardware aspects and is often used to compare different frameworks for parallelism. Therefore, our contributions are a Rust version of NPB, an analysis of the expressiveness and performance of the language features, and parallelization strategies. We compare our implementation with consolidated sequential and parallel versions of NPB. Experimental results show that Rust's sequential version is 1.23\% slower than Fortran and 5.59\% faster than C++, while Rust with Rayon was slower than both Fortran and C++ with OpenMP.

\end{abstract}



%

\IEEEpeerreviewmaketitle



\section{Introduction}
Parallel computing has become a mainstream approach in modern computational systems. With the near end of Moore's Law and the physical constraints on further increasing clock speeds, improvements in single-threaded performance have plateaued. Since multi-core platforms dominate contemporary hardware, developers gradually rely on parallel programming features to effectively tackle complex computational challenges \cite{book_clock}. However, harnessing the power of these architectures introduces significant complexity. Programmers must thoroughly understand the problem and the programming models while navigating obstacles such as data synchronization, load balancing, and communication overhead \cite{book_introduction}. As new frameworks and domain-specific languages for parallelism constantly emerge, selecting the best parallel programming interface can be challenging. Therefore, benchmark applications are crucial for evaluating performance and expressiveness.

Fortran and C/C++ are well-established languages in high-performance computing (HPC). From early message-passing interfaces to modern GPU computing frameworks, they have served as the foundation for developing parallel programming paradigms and abstractions \cite{amaral2020programming, alrawais2021parallel}. Research initiatives tried to improve it with Rust, a modern, promising low-level language designed to offer memory safety and speed. It empowers programmers to write high-level and expressive code without sacrificing performance or fine control over system resources \cite{rustbook}. Memory access complications such as use-after-free, buffer overflows, or data races are common causes of software issues \cite{membug2camb, membug3narure}. As such, Rust stands as an attractive solution for developers. Its ownership-based system allows the compiler to detect data race problems at compile time, ensuring safe memory access without using garbage collection \cite{rustbook}. Considering the growing demand for high performance and Rust's safety guarantees, the HPC community has been investigating its applicability in the area \cite{PIEPER:COLA:21}. Thus, exploring parallelism in Rust remains a prominent open research topic. 

As Rust's ecosystem and parallel capabilities evolve, representative benchmark suites remain essential and will continue to be valuable for the long term. To the best of our knowledge, we have observed a lack of comprehensive benchmarks representing intensive arithmetic tasks and complex scientific applications in Rust (see Section~\ref{sec:rw}). In the wide-ranging landscape of benchmark implementations, the NAS Parallel Benchmarks (NPB) \cite{NPBOriginal1} is among the most renowned suites in the research community. NASA's Numerical Aerodynamic Simulation Program developed the NPB to provide an objective standard for measuring and comparing the performance of highly parallel supercomputers. It includes five kernels (EP, CG, FT, IS, and MG) and three pseudo-applications (BT, SP, and LU), predominantly implemented in Fortran, except for the IS kernel, which was written in C. The suite is available in different versions, including the sequential code and parallel approaches using OpenMP \cite{nasomp} and MPI \cite{NPB-2.0}. Recognizing NPB's relevance, several independent adaptations of the NPB were proposed, including versions for clusters, GPUs, and different programming languages \cite{NPB-CPP-2021, NPB-MPJ, NPB-PY, NPB-CUDA, npb-opencl, npb-openacc}. For instance, the NPB-CPP \cite{NPB-CPP-2021} is a C++-adapted version of the suite. 

We have two goals with this paper: (1) to develop NPB-Rust benchmark so that other developers can test their Rust solutions; and (2) to evaluate Rust performance using NPB-Rust. To that end, we begin by describing the porting methodology used in NPB-Rust's implementation, offering insights for translating code from other languages to Rust. We also provide an in-depth analysis of the challenges of parallel programming in Rust within the NPB context. Since Rayon stands out as the go-to standard for achieving data parallelism in Rust \cite{PIEPER:COLA:21}, we implement it as the parallel programming interface. Then, we perform hypothesis tests to evaluate the performance of sequential and parallel algorithms in NPB-Rust. Finally, 
%
our contributions are:
\begin{enumerate}
    \item An analysis of the Rust language for expressing scientific applications in NPB.
    \item A performance analysis of Rust compared to Fortran and C++ in NPB serial codes.
    \item A performance analysis of Rayon compared to OpenMP in NPB parallel codes.
    \item A new benchmark suite of NPB written in Rust. The source code is available on GitHub\footnote{The source code is available: https://github.com/GMAP/NPB-Rust.}.
\end{enumerate}



Section \ref{sec:back} gives a background in Rust features and outlines the NPB suite proprieties. In section \ref{sec:npb-rust-implementation}, the porting process for both sequential and parallel versions of the NPB-Rust is detailed. Section \ref{sec:expdisc} contains the experiments and results discussion. In Section \ref{sec:rw}, we make an overview of the related work and finalize with the conclusion in Section \ref{sec:conl}.



\section{Background}\label{sec:back}
Since Rust is a relatively new programming language, Section \ref{sec:rust_back} offers a brief overview of its key concepts and innovations. 
Section \ref{sec:par_back} give a background on parallel patterns. 
Section \ref{sec:npb_back} presents the NPB suite's characteristics and details each benchmark.
    
    \subsection{Rust Programming Language}\label{sec:rust_back}
    \textbf{Rust fundamentals.} Rust is a safe and performant low-level language launched by Mozilla as an open-source project \cite{rustbook}. Its safety features derive from an ownership system, where every value has exactly one owner at a time, which is the variable or structure responsible for managing its memory. Once the owner goes out of scope, the value is dropped, avoiding memory leaks and double frees. When a value is assigned to another variable, the ownership is transferred, making the original variable invalid. Rust also allows borrowing, a system where references to data can be passed temporarily without transferring ownership. The borrowing rules permit multiple immutable references to exist simultaneously, allowing shared read-only access, but only one mutable reference can exist at a time to ensure exclusive write access. Rust’s compiler also tracks lifetimes, which ensures that references never outlive the data they point to, further preventing dangling pointers and undefined behavior \cite{rustbook}.
    
    \textbf{Unsafe Rust.}  A program that violates Rust's safety guarantees will not compile. Conversely, a benign program may also fail to compile if the compiler lacks sufficient information to be confident. This conservative approach makes Rust sound but incomplete \cite{rivera2021keeping}. In order to deal with this, programmers can use unsafe Rust. With this extension, developers can create unsafe blocks to bypass some safety checks enforced by Rust's ownership and borrowing system in specific code regions. When using unsafe Rust, the programmer must manually guarantee that the code respects Rust's rules. Misusing it can lead to memory problems, as happens in other unsafe languages like C/C++ \cite{rustbook}.

    \textbf{Rayon library.} Rayon is a high-level framework for data parallelism that extends Rust's iterator-based traits to support regular parallel operations. For instance, a sequential iterator \texttt{into\_iter()} can be conveniently transformed to parallel by changing it to \texttt{into\_par\_iter()}. Rayon adheres to Rust's safety standards by limiting write access to shared data within parallel iterators. It uses a global thread pool to manage concurrency, maintains multiple threads waiting for tasks to be allocated, and uses a work-stealing system \cite{rayon}. The NPB suite mainly contains data-parallelism computations. Analogous to the OpenMP for Fortran and C++, Rayon can express operations such as \texttt{Map} and \texttt{MapReduce} \cite{map}, which are the main parallel patterns in the NPB. This paper uses Rayon to create a parallel version of the NPB-Rust.

        \begin{figure}[t]
        \centering
        \includegraphics[width=0.36\textwidth]{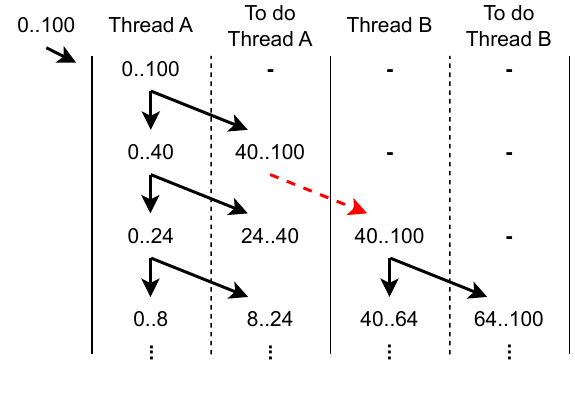}
        \vspace{-0.4cm}
        \caption{Rayon's work stealing methodology.}
        \vspace{-0.4cm}
        \label{fig:ws}
        \end{figure}

         \textbf{Work-stealing.} Rayon implements a work-stealing load-balancing strategy to manage threads' dynamically. A thread that receives an iterative task splits it into two parts: one to process immediately and another to defer for later execution. Idle threads in a thread pool continuously search for a job, and eventually, they check the to-do list of other threads. Figure \ref{fig:ws} illustrates this technique, where the idle thread \textit{B} steals a task from the to-do list of thread \textit{A}.

    \begin{figure}[!b!]
        \vspace{-0.8cm}
         \centering
         \subfloat[Sequential \texttt{Map}]{\includegraphics[width=0.16\textwidth]{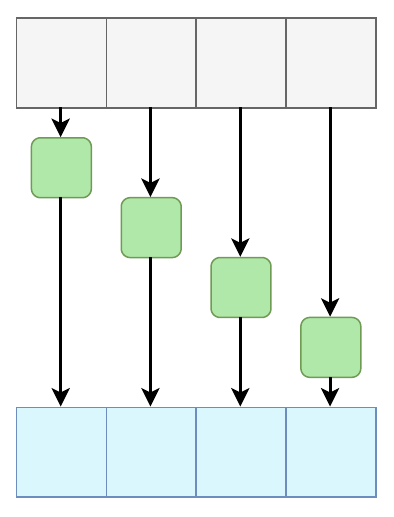}\label{fig:map_ser}}
         \subfloat[Parallel \texttt{Map}]{\includegraphics[width=0.16\textwidth]{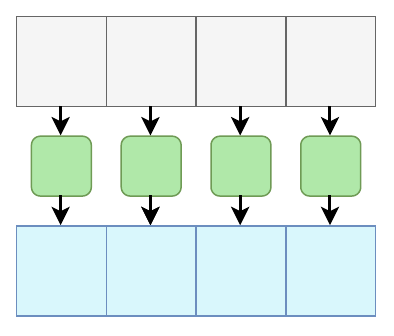}\label{fig:map_par}}
         \subfloat[Parallel \texttt{MapReduce}]{\includegraphics[width=0.16\textwidth]{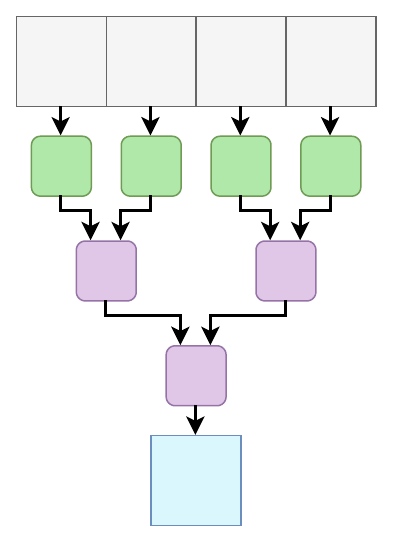}\label{fig:map_reduce}}
         \caption{Illustration of sequential \texttt{Map}, parallel \texttt{Map}, and parallel \texttt{MapReduce}.}
         \vspace{0.8cm}

    \lstset{aboveskip=0pt, belowskip=7pt}
    \begin{lstlisting}[language=C++, style=boxed,caption={\texttt{Map} and \texttt{MapReduce} with OpenMP in C++.}, label={lst:map_c}]
std::vector<int> results(n);
#pragma omp parallel for
for(i=0; i<n; i++){
  results[i] = /* computation */
}

int sum = 0;
#pragma omp parallel for reduction (+:sum)
for(i=0; i<n; i++){
  sum += /* computation */
}
\end{lstlisting}
\vspace{0.4cm}
    \lstset{aboveskip=0pt, belowskip=0pt}
    \begin{lstlisting}[language=Rust, style=boxed,caption={\texttt{Map} and \texttt{MapReduce} with Rayon in Rust.}, label={lst:map_r}]
let results: Vec<_> = (0..n).into_par_iter()
  .map(|i| {
    /* computation */
  })
.collect();

let sum: i32 = (0..n).into_par_iter()
  .map(|i| {
    /* computation */
  })
.reduce(|| 0, |acc, x| acc + x);
\end{lstlisting}
    \end{figure}

    \subsection{Parallel Patterns in NPB}\label{sec:par_back}
    Parallel patterns can be understood in two parts: their high-level semantics and their low-level implementation details. High-level semantics refers to when and where a pattern can be applied in sequential code and any limitations to its use. The low-level implementation deals with the technical details hidden from the programmer, like communication methods, synchronization, and task scheduling. By focusing on the high-level purpose, programmers can add parallelism to their code without dealing with the complex details of how it works. Each parallel programming framework provides its own version of these patterns with unique interfaces and internal designs. As a result, patterns may differ in usability and efficiency based on their design and implementation \cite{NPB-CPP-2021}. Rayon and OpenMP provide different APIs to express data-parallelism, but both can be described using the abstract \texttt{Map} and \texttt{MapReduce}.

    \textbf{Map.} The \texttt{Map} pattern \cite{map} involves applying a set of identical operations to all elements in a collection. This can be used to parallelize loops without data dependencies between iterations. Figures \ref{fig:map_ser} and \ref{fig:map_par} illustrate how the sequential and parallel \texttt{Map} pattern works. The white squares represent the initial values, the blue squares represent the final values, and the green square represents the operation. In OpenMP, programmers annotate parallel \texttt{for} loops using compiler preprocessor directives. Lines 1 to 5 in Listing \ref{lst:map_c} showcase how to implement a parallel \texttt{Map} in C++ with OpenMP. Rayon expresses parallelism by extending the iterator traits to parallel versions, lines 1 to 5 in Listing \ref{lst:map_r} demonstrates how to express the parallel \texttt{Map} pattern in Rust.

    \textbf{MapReduce.} The \texttt{MapReduce} pattern \cite{map} is the union of the \texttt{Map} with the \texttt{Reduce} operation. In the reduction process, the elements from a collection are combined into a single output. Figure \ref{fig:map_reduce} illustrates how the parallel \texttt{MapReduce} works, where the purple squares represent the reduction operation. This pattern can be used to parallelize \texttt{for} loops that exhibit specific data dependencies, and synchronization is necessary. In Listing \ref{lst:map_c}, lines 7 to 11 indicate how to apply this pattern in C++ using the OpenMP annotation. Lines 7 to 11 in Listing \ref{lst:map_r} show a manual implementation of a reduction function applied after a \texttt{Map} operation, leveraging iterator-based traits.

    \textbf{Barrier.} By default, in the end of parallel \texttt{Map} and \texttt{MapReduce} regions, there are an implicit barrier. It is a synchronization primitive for a group of threads \cite{map}. The barrier is a point where any thread must stop and can not proceed until all other threads reach it. With the \texttt{nowait} directive is possible to avoid this synchronization in OpenMP, on the other hand, the Rayon library does not provide a similar approach. It is also possible to add explicit barriers in the code, this feature is available for Fortran, C++, and Rust.

    \subsection{NAS Parallel Benchmarks}\label{sec:npb_back}
    Originally stemming from the field of computational fluid dynamics (CFD), the NPB approximates real-world workloads and mathematical methods. The suite's applications encompass various computational characteristics, including irregular memory access patterns, complex data dependencies, and intensive data communication requirements \cite{NPBOriginal1}. 
    
    Each application offers different levels of problem sizes, which are configurable through the selection of a specific class at compile time. Classes S and W represent the smallest scale and are intended for conducting small tests. The standard test classes are A, B, and C, where the complexity increases fourfold with each ascending class. Finally, Classes D, E, and F are specifically designed for highly intensive tests, with the complexity escalating sixteen-fold with each step up in class.
    
\begin{figure*}[h!t!]
        \centering
        \includegraphics[width=1\textwidth]{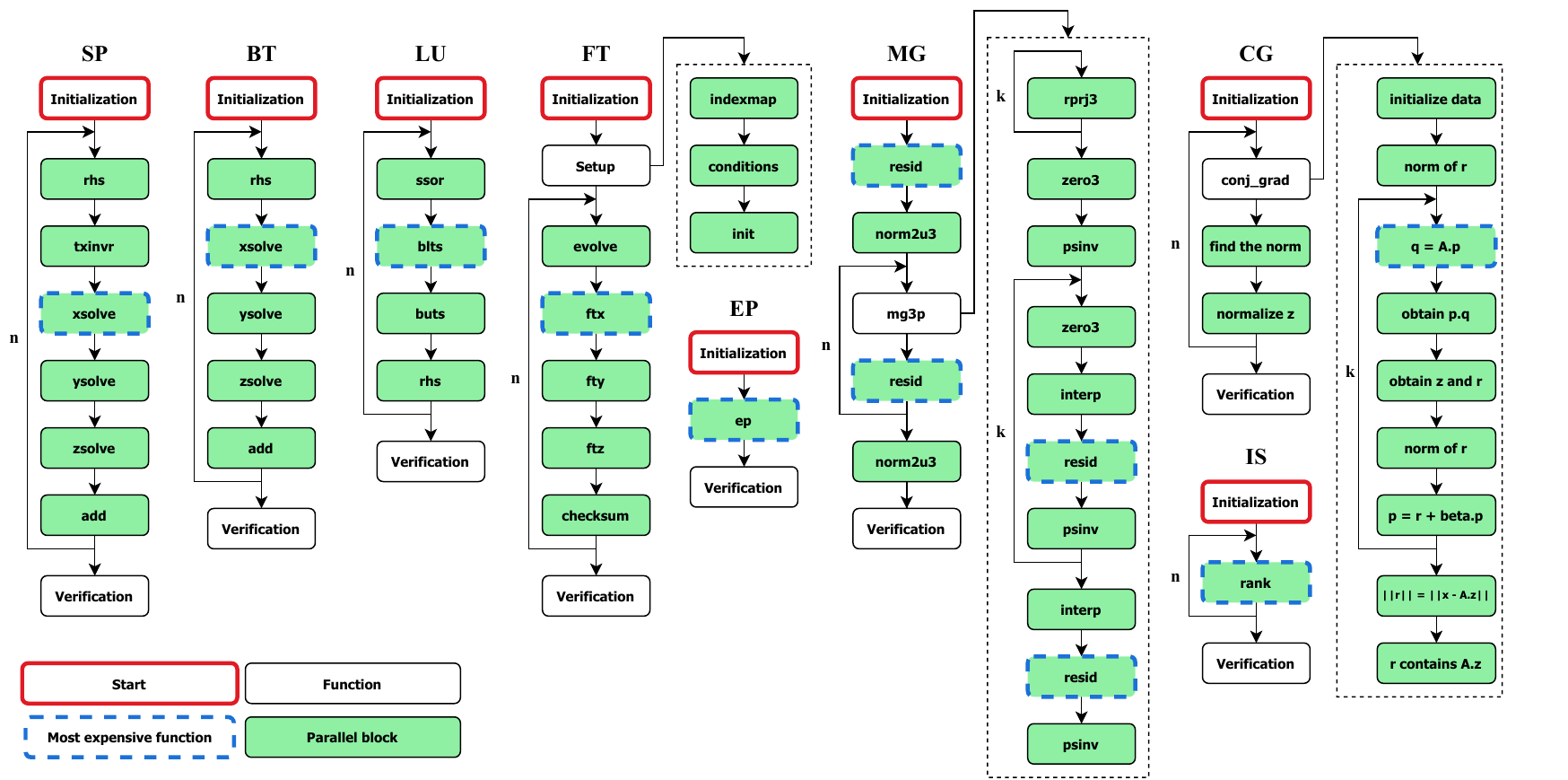}
        \caption{NPB kernels and pseudo-applications flowchart. Adapted from \cite{beyond}.}
        \label{fig:npb-flowchart}
        \vspace{-0.4cm}
        \end{figure*}

    The eight benchmarks starts with an initialization phase and concludes with a thoroughly built-in verification mechanism to check the results' correctness. The applications feature automated measurement of execution time and performance estimation in terms of mega floating-point operations per second (MFLOPS). The five kernels and three pseudo-applications are described as follows, and Figure \ref{fig:npb-flowchart} presents a detailed flowchart of the NPB programs.

        \textbf{Embarrassingly Parallel (EP).}
        This kernel primarily consists of two key processes. It generates a large set of Gaussian pseudo-random numbers and then accumulate two-dimensional statistics from those numbers. While this problem can be readily divided into concurrent components, its low communication overhead makes it suitable for estimating the floating-point operational capacity of the target hardware \cite{npb-results-11-96}. 

        \textbf{Conjugate Gradient (CG).}
        Employs an iterative conjugate gradient algorithm to approximate the smallest eigenvalue of a large and unstructured matrix. Induces long-distance communication and performs sparse matrix-vector multiplications. Due to the irregular nature of the computations, this kernel also imposes demands on memory locality \cite{npb-results-11-96}.
        
        \textbf{Discrete 3D Fast Fourier Transform (FT).}
        Performs a Fast Fourier Transform in three-dimensional matrices of complex numbers to solve a three-dimensional partial differential equation. Apply both spectral and inverse methods within nested loops. The kernel is specifically designed to test exhaustive long-distance communication performance.\cite{npb-results-11-96, NPB-2.0}.

        \textbf{Integer Sort (IS).}
        A bucket-sorting algorithm implementation that operates over a sparse set of integer numbers. This kernel focuses on testing integer computation speed and communication performance without involving any intensive floating-point operations. It simulates a computational task analogous to particle-in-cell applications in physics. \cite{npb-results-11-96, nasomp}.

        \textbf{Multi Grid (MG).}
        A simplified multi-grid V-cycle operation and residual calculation using constant coefficients instead of variable ones. It is designed to solve a three-dimensional scalar Poisson equation. The benchmark is intended for performing high-volume structured short- and long-distance communications \cite{npb-results-11-96, NPB-2.0}.

        \textbf{Block Tri-diagonal Solver (BT).}
        Simulates a CFD application that solves three-dimensional compressible Navier-Stokes equations. The finite differences solution is based on an Alternating Direction Implicit (ADI) approximate factorization of the dimensions. Produces block-tridiagonal systems that solve the unknown vectors using back substitution \cite{npb-results-11-96, NPB-2.0}.

        \textbf{Scalar Penta-diagonal Solver (SP).}
        This pseudo-application is similar to BT. It is a simulated CFD program that solves multiple independent systems of scalar penta-diagonal equations, which are non-diagonally dominant. The finite differences solution is based on a Beam-Warming approximate factorization of the dimensions \cite{npb-results-11-96, NPB-2.0}.

        \textbf{Lower-Upper Gauss-Seidel Solver (LU).}
        Represents the computations associated with a class of CFD algorithms. This pseudo-application employs a symmetric successive over-relaxation (SSOR) numerical scheme to solve a regular-sparse, lower and upper triangular system. It is a variation of the Gauss-Seidel method for linear system equations \cite{npb-results-11-96, NPB-2.0}.



\section{Rust NPB Implementation}
\label{sec:npb-rust-implementation}

Section \ref{sec:sub:npb-rust-seq} outlines the challenges encountered when porting NPB to Rust, detailing the strategies to reach efficient performance in Rust. Section \ref{sec:sub:npb-rust-par} presents the methods used to achieve parallelism in each kernel and pseudo-application, highlighting the optimizations made to leverage Rust's parallel features, and overview the unsafe code usage on the NPB-Rust.

    \subsection{NPB-Rust Sequential Porting}
    \label{sec:sub:npb-rust-seq}
    
    \textbf{Design principles.} The sequential porting of the NPB to Rust was based on NPB-CPP \cite{NPB-CPP-2021}. We have two main design principles: (1) retaining algorithmic structure, preserving naming conventions and following the original sequence of operations; (2) idiomatic Rust, ensuring that the port not only mirrors the original in functionality but also embodies the best practices and safety of the Rust language.

    \textbf{Project.} The NPB-Rust project was developed with Cargo, the Rust package manager. Each benchmark was treated as a distinct application that generates a separate binary. Common auxiliary programs, such as the \texttt{timers}, \texttt{randdp}, and \texttt{print\_results}, were also ported. Since Rust does not have a built-in variable type for complex numbers (used in the FT kernel), a custom implementation was added to the common files, similar to the C++ version. Regarding the class parameters, each benchmark now includes the full definition of the different problem sizes within its own structure. Class selection still occurs at compile time through a mandatory \texttt{RUSTFLAGS} specification and with the usage of Rust's conditional configuration feature. 

    \textbf{Porting process.} Converting the NPB to Rust is challenging and demands significant effort. The main obstacles involve dealing with global mutable variables, keeping coherence in memory allocation, function calls and operations that violate the borrowing rules, and converting \texttt{for} loops to iterator-based constructs. Rust does not provide a safe and mutable global variable approach without using locks. Consequently, every time a global variable is declared in C++, in Rust, the equivalent variable is declared inside the main function and passed as a mutable reference parameter whenever needed. Similarly, all the \texttt{\#define} preprocessor directives and the \texttt{static const} variables from the pseudo-applications were replaced by Rust's \texttt{pub const} notation. 

    In general, static memory allocation offers better performance than dynamic allocation. However, when dealing with very large arrays, static allocation may lead to stack overflow errors. Following the recommendations from the official NPB reports and to ensure a fair comparison across the different NPB implementations, arrays that were originally allocated using \texttt{malloc} in C++ were replaced with \texttt{Vec} in Rust, which is heap-allocated. For other cases, static memory allocation using arrays was employed to maintain efficiency.

    In rare instances within the C++ code, a specific function is invoked with the same mutable variable passed as two different parameters, while in other cases, distinct mutable variables are needed. Since this violates Rust's borrowing rules, we created a duplicate of the function with modified receiving parameters to accommodate both situations. This solution effectively manages cases where either the same variable is used twice or two distinct variables are passed.
 
    To guarantee a well-executed porting process, it is crucial to adapt the loop structures appropriately. The NPB-CPP source code predominantly uses \texttt{for} statements to manipulate indices and pointers. In Rust, the idiomatic strategy involves leveraging iterator-based patterns. However, it is important to note that not all \texttt{for} statements can be seamlessly converted to iterators. Complex loops with intricate logic or indirect index calculation may not be feasible to transform organically into iterator-based constructs. 

    The main obstacle during the porting phase was the MG kernel. In the original C++ code, arrays are defined as single-dimensional entities with their maximum length fixed at compile time. Still, during execution, these arrays are dynamically reshaped into three-dimensional structures within function calls, utilizing void pointers. The dimensions of these arrays are redefined at runtime, with their lengths varying as execution progresses. In Rust, those procedures lead to unsafe operations. To avoid this problem, we preserved the arrays as one-dimensional and applied arithmetic to compute the appropriate indices, simulating a three-dimensional array.

   The NPB and NPB-CPP applications are designed to fully utilize the available resources to maximize performance. The pursuit of greater speedup has led us to explore unsafe Rust regions. We developed a serial version of the FT, IS, and MG kernels that incorporates unsafe code, while all other benchmarks are entirely safe Rust. The unsafe operations were introduced to bypass bound checks in a few performance hot spots. Lastly,  by using the \texttt{--cfg safe="true"} option in \texttt{RUSTFLAGS}, it is possible to compile a safe sequential version of these kernels.

    \subsection{NPB-Rust with Rayon}
    \label{sec:sub:npb-rust-par}
        \textbf{EP kernel.}
        The EP kernel is one of the simpler NPB programs in terms of parallelism, it features a single highly intensive computational region that calculates the independent total sum of Gaussian deviations. The partial count values of each iteration are also accumulated in a vector. Following a similar strategy to the C++ parallel implementation, we applied a \texttt{MapReduce} pattern to this region to efficiently sum and accumulate the values. The main difference compared to the OpenMP version is that, instead of using a mutual exclusion section to manage the accumulation in the vector, we update it directly within Rayon's reduction lambda function.
        
        \textbf{CG kernel.}
        The central component of this kernel revolves around the \texttt{conj\_grad} function, which performs the computationally expensive sparse matrix-vector multiplications. This function contains a sequence of parallel regions where we incorporated the \texttt{Map} and \texttt{MapReduce} patterns. Additionally, the kernel includes two concurrent sections related to post-conjugate gradient normalization, where the same patterns were applied. Finally, the \texttt{Map} was utilized to optimize the untimed initialization phases. The major contrast with C++ in this kernel is that, in the OpenMP version, \texttt{nowait} directives were used to optimize the \texttt{parallel for} intrinsic barriers inside the \texttt{conj\_grad} function.

        \textbf{FT kernel.}
        The core computational routines of this kernel perform independent symmetric Fast Fourier Transforms across three dimensions. These functions were parallelized using the \texttt{Map} pattern. However, extra care was required to ensure performance and safety for the \texttt{cffts3} and \texttt{cffts3\_2} functions. The main loops in these routines iterate over the second dimension in the outermost \texttt{for} loop, while the first dimension is processed in an inner loop. Since it is not possible to create an iterator that processes the second dimension before the first, and the intricate logic restricts loop reordering, we relied on unsafe code. Listing \ref{lst:FT_par_exe} shows how an unsafe mutable pointer was used to access a mutable reference to an external variable inside the parallel iterator. This approach is safe because the variable used as an index is always unique and thread-private. Other parallel regions in the FT kernel occur in the initialization steps and within the \texttt{evolve} function, where the \texttt{Map} pattern was also applied. Finally, for the checksum calculation over complex numbers, the \texttt{MapReduce} pattern was employed.

\begin{figure}[t]
\lstset{aboveskip=0pt, belowskip=-17pt}
\begin{lstlisting}[language=Rust, style=boxed,caption={Simplified representation of the unsafe array access on \texttt{cffts3} and \texttt{cffts3\_2} functions in the FT kernel.}, label={lst:FT_par_exe}]
pub struct UnsPtr(pub *mut [[f64; NX]; NY]);
unsafe impl Sync for UnsPtr {}

let my_array = /* some three-dimensional data */;
let ptr = UnsPtr(my_array.as_mut_ptr());

(0..NY).into_par_iter().for_each(|j| {
  let my_array = unsafe {
    &mut from_raw_parts_mut((&ptr).0, NZ)[..]
  };
  
  /* computation indexing my_array with 
     thread-private and unique j variable */
     
});
\end{lstlisting}
\end{figure}


        \textbf{IS kernel.}
        The IS kernel is centered around a primary function that implements a bucket sort algorithm, supported by auxiliary functions for key sequence generation and result validation. In the OpenMP version, thread identifiers serve as indices for statically partitioning the workload across threads within parallel regions. The Rust implementation with Rayon mirrors this structure, but in some cases, this indirect index manipulation led to the use of unsafe code. Similarly to Listing \ref{lst:FT_par_exe}, it makes the primary data array a mutable reference. The safety of this implementation is assured, as the indexing relies on manually calculated values that depend on the thread identifier. The \texttt{MapReduce} pattern was applied in the verification function to count incorrectly sorted values, while the \texttt{Map} pattern was applied in the other regions.



        \textbf{MG kernel.}
        This kernel integrates multiple functions featuring parallel regions. The principal computationally intensive routines that operate over the V-cycle are the restriction, prolongation, residual, and smoother functions. Each of these methods is structured around a main loop, where the \texttt{Map} were employed to achieve parallelism. Other small computations were also parallelized with this same pattern. The \texttt{MapReduce} was only used for the uniform norm calculus. Due to the sequential porting strategy, the parallel MG kernel necessitates the use of unsafe blocks. In Rust, one-dimensional arrays were chosen over three-dimensional arrays due to the need for frequent reshaping during execution, making it essential to apply arithmetic operations to compute the correct indices. This approach restricts the ability to fully utilize parallel iterator-based constructs without resorting to unsafe code, similar to Listing \ref{lst:FT_par_exe}. However, despite the need for unsafe blocks, thread safety is maintained. Each thread retains private control over index values, preventing memory races and preserving the kernel's integrity.


        \begin{figure}[t]
        
        \centering
        \includegraphics[width=0.25\textwidth]{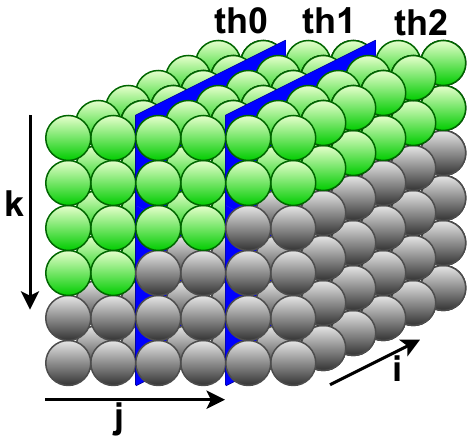}
        \caption{LU data dependencies illustration.}
        \label{fig:ludaata}
        \vspace{-0.5cm}
        \end{figure}

        \textbf{BT pseudo-application.}
        The BT pseudo-application employs iterative techniques to solve CFD equations and was designed to achieve parallelism. The most computationally demanding operations include residual calculations, and solvers applied across all three spatial dimensions. With data primarily structured along the \textit{z}-dimension, parallelism was achieved in the \textit{x} and \textit{y} dimensions by applying the \texttt{Map} pattern in the outermost loop for residual and solver computations. However, due to data dependencies along the \textit{z}-dimension, the \texttt{Map} pattern was applied within the inner loops to compute residuals. Efficient parallelization of the \textit{z}-solver required the use of unsafe code. As in the FT kernel, this routine iterates over the vector's second dimension in the outer \texttt{for} loop, where a significant volume of inner computation prevents efficient loop reordering. Thus, the same approach was applied in FT. Safety is ensured because a unique, thread-private value is used as an index to write on the vector.

        \textbf{SP pseudo-application.}
        This pseudo-application is also an iterative method derived from CFD. While the specific procedures differ, the code follows a similar pattern to BT, involving residual and solver calculations for each spatial dimension. As a result, the same challenges are encountered. Parallelization across the three dimensions was achieved using the \texttt{Map} pattern. Particular care was required for the \textit{z}-dimension, where the same approach as in BT was applied. For SP, we also increased the thread pool stack size. This was necessary to keep coherence with the C++ version, which uses the stack to allocate large data inside solver functions.

        \textbf{LU pseudo-application.}
        The LU benchmark is the most complex program in the NPB suite in terms of parallelism. Efficient parallelization is challenging due to data dependencies across all spatial dimensions in \texttt{blts} and \texttt{buts} functions. To introduce parallelism, the pipeline technique was employed, utilizing a synchronization mechanism to progressively grant ordered access to data \cite{nasomp}. We implemented locks and conditional variables, following a similar approach used in the parallel version of NPB-CPP \cite{NPB-CPP-2021} with FastFlow and OneTBB frameworks. As illustrated in Figure \ref{fig:ludaata}, the computation is oriented along the \textit{k}-dimension. Once a thread completes its task, the subsequent thread is allowed to proceed. The green spheres represent the computed data, the gray spheres indicate data yet to be processed, and the blue plane represents the division of data among the threads.

    \newcolumntype{C}[1]{>{\centering\arraybackslash}p{#1}}
    \newcolumntype{M}[1]{>{\centering\arraybackslash}m{#1}}
    \begin{table}[t]
    \scriptsize
    \centering
    \vspace{0.2cm}
    \caption{Unsafe blocks usage in NPB-Rust.}
    \begin{tabular}{C{1.2cm}|C{0.4cm}C{0.4cm}C{0.4cm}C{0.4cm}C{0.4cm}C{0.4cm}C{0.4cm}C{0.4cm}} 
        \toprule
    Purpose & \multicolumn{8}{c}{Benchmarks} \\
    & EP & CG & FT & IS & MG & BT & SP & LU \\
    \midrule
        1) & 0 & 0 & 9 & 3 & 8 & 0 & 0 & 0  \\
        2) & 0 & 0 & 2 & 0 & 0 & 1 & 1 & 2  \\
        3) & 0 & 0 & 0 & 5 & 10 & 0 & 0 & 1 \\
        \bottomrule
    \end{tabular}
    \vspace{0.3cm}
    \begin{enumerate}
        \item Avoid bound checks;
        \item Modify an external array inside a parallel iterator due to the non-sequential iteration order through array dimensions;
        \item Modify an external array inside a parallel iterator due to the intricate logic to find/calculate the index.
    \end{enumerate}
    \label{tab:unsafe}
    \vspace{-0.2cm}
    \end{table}
    
    \textbf{Unsafe code usage.} The sequential version of NPB-Rust includes instances of unsafe code in the FT, IS, and MG kernels, exclusively to bypass bounds checks. In the parallel implementations using Rayon, unsafe features were applied to manage mutable shared data within parallel iterators, where safety arose due to indirect factors. These cases typically involved iterating over dimensions in a non-sequential order, where data dependencies or computational constraints made loop reordering infeasible. In some cases, unsafe code was required for accessing data through indirect index manipulation, where the index calculation relied on thread-private values. Table \ref{tab:unsafe} indicates the unsafe blocks applied in the NPB-Rust.

\section{Experiments and Discussion}\label{sec:expdisc}
The experiments were carried out in a multicore platform with the following architectural configuration: a dual-socket Intel Xeon Silver 4210 clocked at 2.2GHz, with ten cores per socket totaling 20 cores (40 with Hyper-Threading). Regarding memory, it contains 640KiB of private L1, 20MiB of private L2, 27.5MiB shared L3, and 148GB of RAM. The system is a Linux Ubuntu 20.04.6 LTS. The following compilers were utilized for Rust, C++, and Fortran, respectively: \texttt{rustc} 1.81.0, \texttt{clang} 10.0.0 and \texttt{gfortran} 9.4. The reliability of the results uses the benchmark-specified maximum error range compared to the stored reference values. 

Our set of experiments was guided by the following question: \textit{Does the Rust language provide sufficient performance and expressiveness for the NPB algorithms?} To answer this question we test two hypotheses: 1) The Rust language enables the expression of sequential algorithms for NPB with superior performance to Fortran and C++; and 2) The Rayon framework enables the expression of parallel algorithms for NPB with equal performance to OpenMP. In Section \ref{sec:seq} we present the the sequential version analyses, including a comparison between the unsafe and safe versions of NPB-Rust. Finally, in Section \ref{sec:par} we compare the parallel implementations in terms of execution time, scalability, speedup, memory consumption, and programmability. 

All experiments use the NPB class C, a typical workload for better granularity on multi-core platforms. Each test was executed 10 times, and the reported results represent the arithmetic mean and the standard deviation. We also leverage statistical tests using a 95\% confidence level to compare the differences in the execution times for sequential and parallel versions, following a previously applied methodology \cite{NPB-CPP-2021}. For compiling the benchmarks in Fortran and C++, we specified \texttt{-std=c++20} and \texttt{-O3}. In Rust we used the \texttt{--release}, this flag enables high-level optimizations, similar to the \texttt{-O3} option in Fortran and C++.



    \subsection{Sequential Version}\label{sec:seq}

    \begin{figure}[t]
        \centering
        \includegraphics[width=0.5\textwidth]{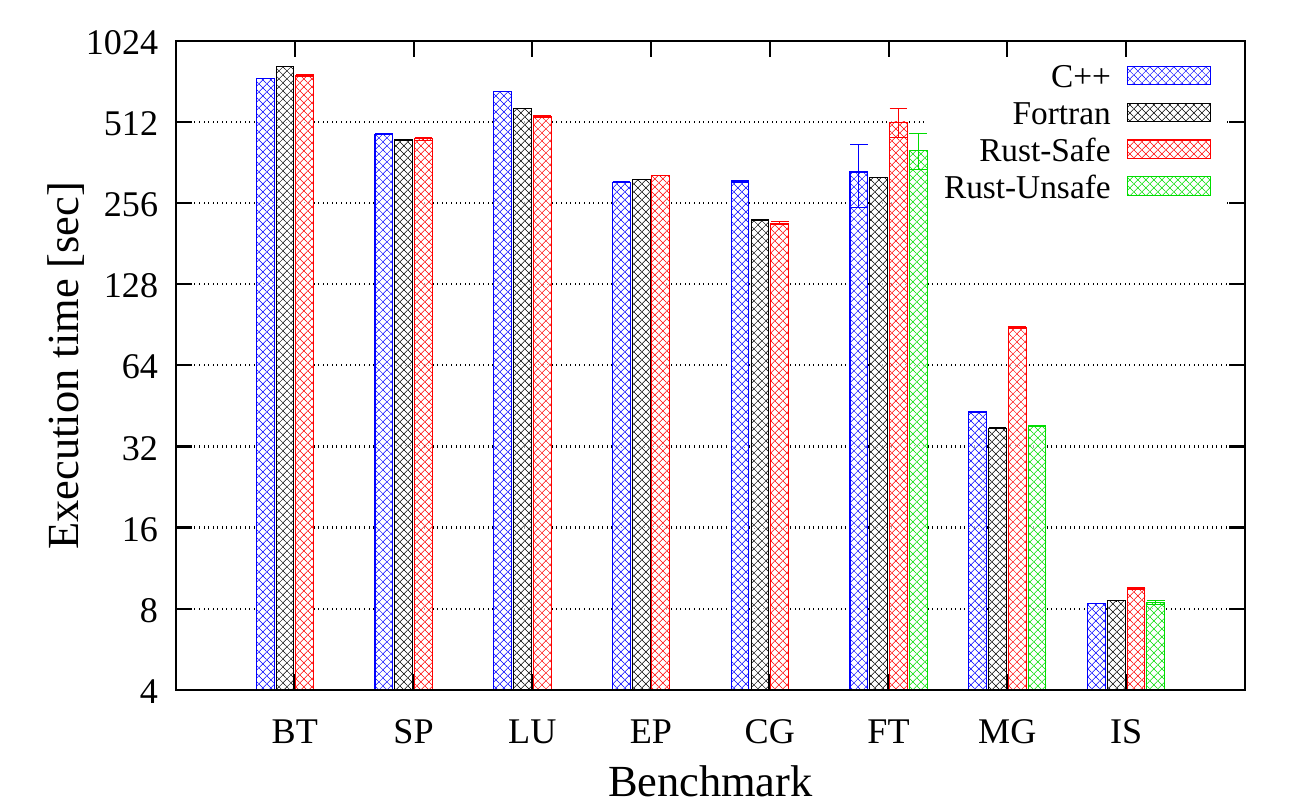}
        \caption{Sequential versions execution time.}
        \label{fig:sertimes}
        \end{figure}

        \begin{table}[t]
        \scriptsize
        \centering
        \caption{Sequential relative differences in execution time.}
        \begin{tabular}{C{1.5cm}|C{2.5cm}C{2.5cm}} 
            \toprule
            Benchmark & Rust to Fortran & Rust to C++ \\ 
            \midrule
            EP  & \cellcolor{pastelred} + 3.23\% & \cellcolor{pastelred} + 5.77\%  \\
            CG  & \cellcolor{pastelgreen} - 2.92\% & \cellcolor{pastelgreen} - 29.92\%  \\
            FT  & \cellcolor{pastelred} + 18.18\% & \cellcolor{pastelred} + 12.93 \%  \\
            IS  & \cellcolor{pastelgreen} - 1.62\% & \cellcolor{pastelred} + 4.57e-5\%  \\
            MG  & \cellcolor{pastelred} + 1,67\% & \cellcolor{pastelgreen} - 11.54\%  \\
            BT  & \cellcolor{pastelgreen} - 7.31\% & \cellcolor{pastelred} + 2.44\%  \\
            SP  & \cellcolor{pastelred} + 0.75\% & \cellcolor{pastelgreen} - 3.91\%  \\
            LU  & \cellcolor{pastelgreen} - 6.54\% & \cellcolor{pastelgreen} - 19.21\%  \\
            \bottomrule
        \end{tabular}
        \label{tab:relativeseq}
        \end{table}

    Figure \ref{fig:sertimes} compares the sequential NPB, NPB-CPP, and NPB-Rust versions. The X-axis lists the benchmarks, while the Y-axis represents execution time in seconds on a logarithmic scale. To statistically determine whether the outputs of each NPB-Rust benchmark were equivalent to those of the NPB in Fortran and C++, we consider the \texttt{p-value} obtained on a hypothesis test. To that end, we first have to understand the distribution of each sample to be compared. We ran a Shapiro–Wilk homogeneity test, which is used to identify which kind of hypothesis test to perform (parametric or non-parametric), even with a small sample (10 executions). If both samples were normally distributed, we applied a paired T-test; otherwise, we used the Wilcoxon test. In this analysis, the null hypothesis ($H_0$) is that the outputs from the different implementations are equivalent, while the alternative hypothesis ($H_1$) is that they are significantly different. If the resulting \texttt{p-value} from the statistical test is less than 0.05 (commonly used threshold in the literature for software experiments) we reject $H_0$ with a 95\% confidence level. Since almost all executions in our experiments set have a very low standard deviation on execution time, we fail to reject the $H_0$ hypothesis only when comparing the IS from Rust to C++.


    Table \ref{tab:relativeseq} illustrates the Rust relative differences in execution time to Fortran and C++. A negative value indicates that Rust is faster (green), while a positive value signifies that Rust is slower (red). The results from the unsafe Rust implementation were used. The most considerable performance differences were observed in the CG, FT, MG, and LU kernels, each varying by over 10\% in at least one implementation compared to Rust. The slower performance of the LU and MG kernels in C++ when using the \texttt{clang} compiler was previously reported \cite{NPB-CPP-2021}. Fortran outperformed C++ and Rust for the FT kernel, potentially due to differences in the C++ porting process that were reflected in the Rust version. Fortran also benefits from an intrinsic complex number type, absent in C++ and Rust. Additionally, FT showed a higher standard deviation in execution times for C++ and Rust. The CG kernel in C++ had a longer execution time than in Fortran and Rust. Further analysis showed that, while cache misses were comparable, the C++ version of CG had a higher number of cache references, suggesting increased I/O activity.


    The sequential unsafe version of IS has a total of 3 unsafe blocks, while the FT kernel has 9 unsafe regions. When comparing the performance gain on execution time we observed a difference of 11.2\% on IS and 21.35\% on FT, compared to the safe version. In MG, due to the sequential porting, almost all array access in the computationally intensive functions is based on arithmetic calculations, which results in excessive bound checks. When removing it, we achieved a difference of 56.94\% compared to the safe version.
    The statistical test shows that the results are majority not equivalent between Rust and the other versions. Also, the geometric mean of the execution times indicates that Rust is 1.23\% slower than Fortran and 5.59\% faster than C++ in our experiments on the NPB suite. We then concluded that our hypothesis that Rust enables the expression of sequential algorithms for NPB with superior performance to Fortran and C++ is partially correct.




    \begin{figure*}[t]
         \centering
          \vspace{0.2cm}
         \subfloat[EP]{\includegraphics[width=0.33\textwidth]{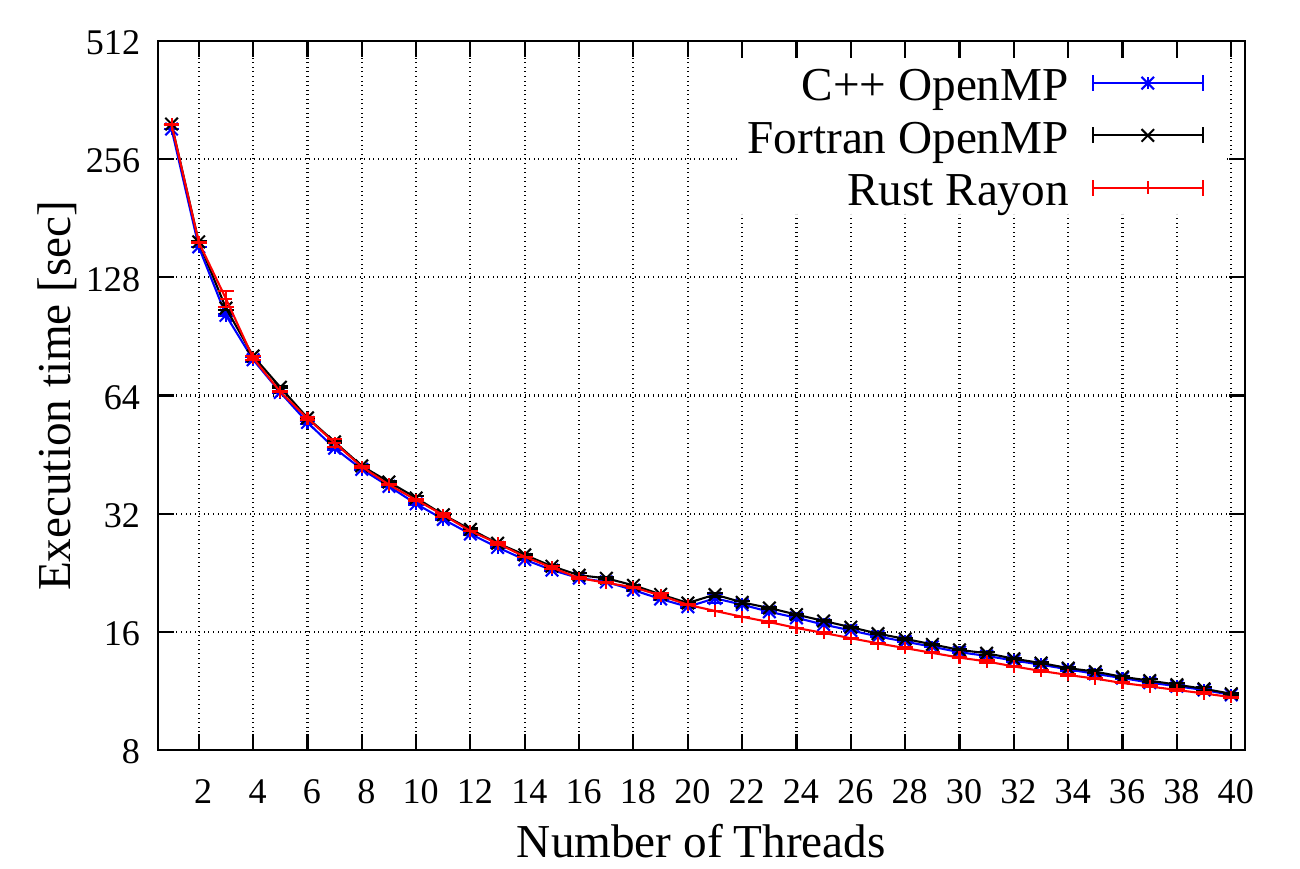}\label{fig:fortranvscpp_ep_b}}
         \subfloat[CG]{\includegraphics[width=0.33\textwidth]{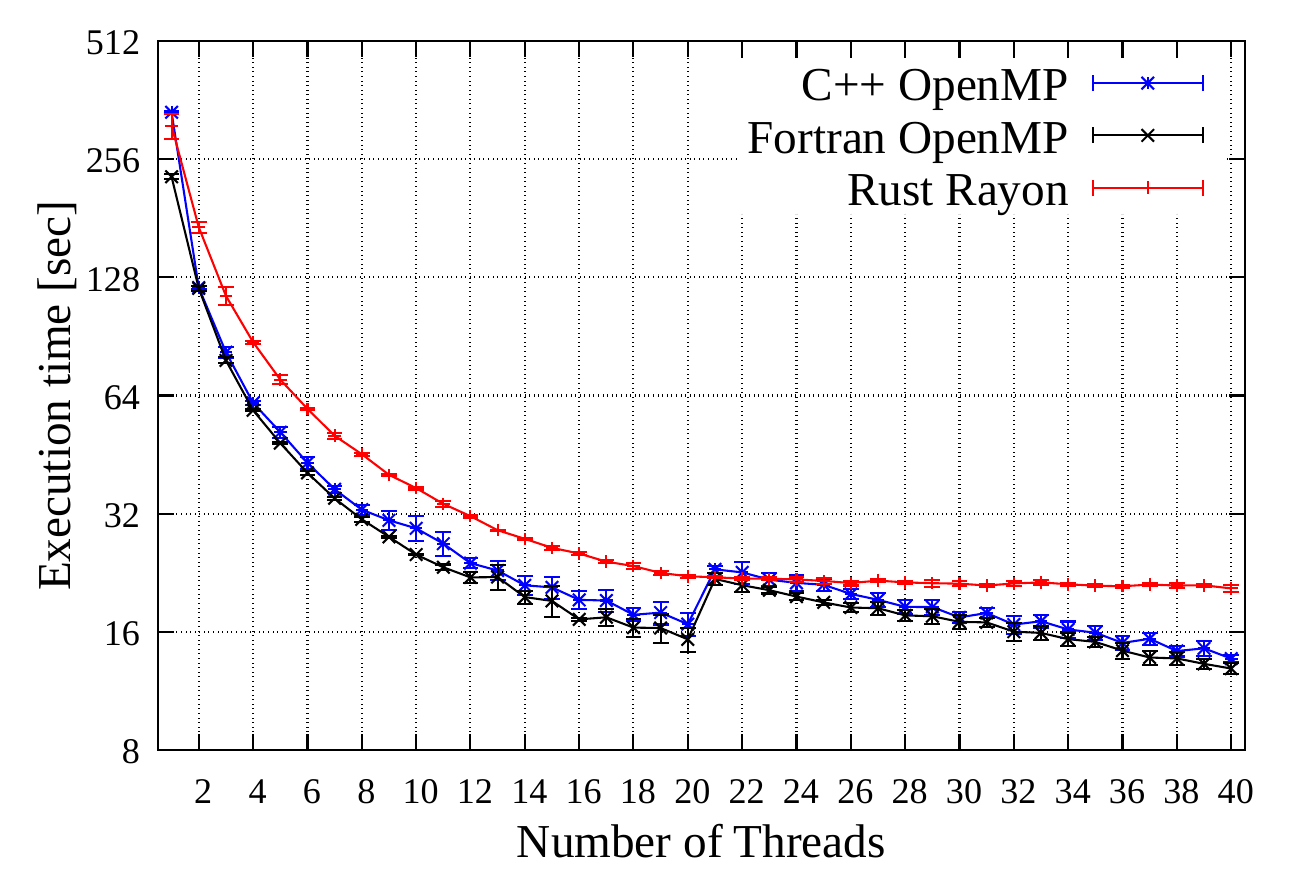}\label{fig:fortranvscpp_cg_b}}
         \subfloat[FT]{\includegraphics[width=0.33\textwidth]{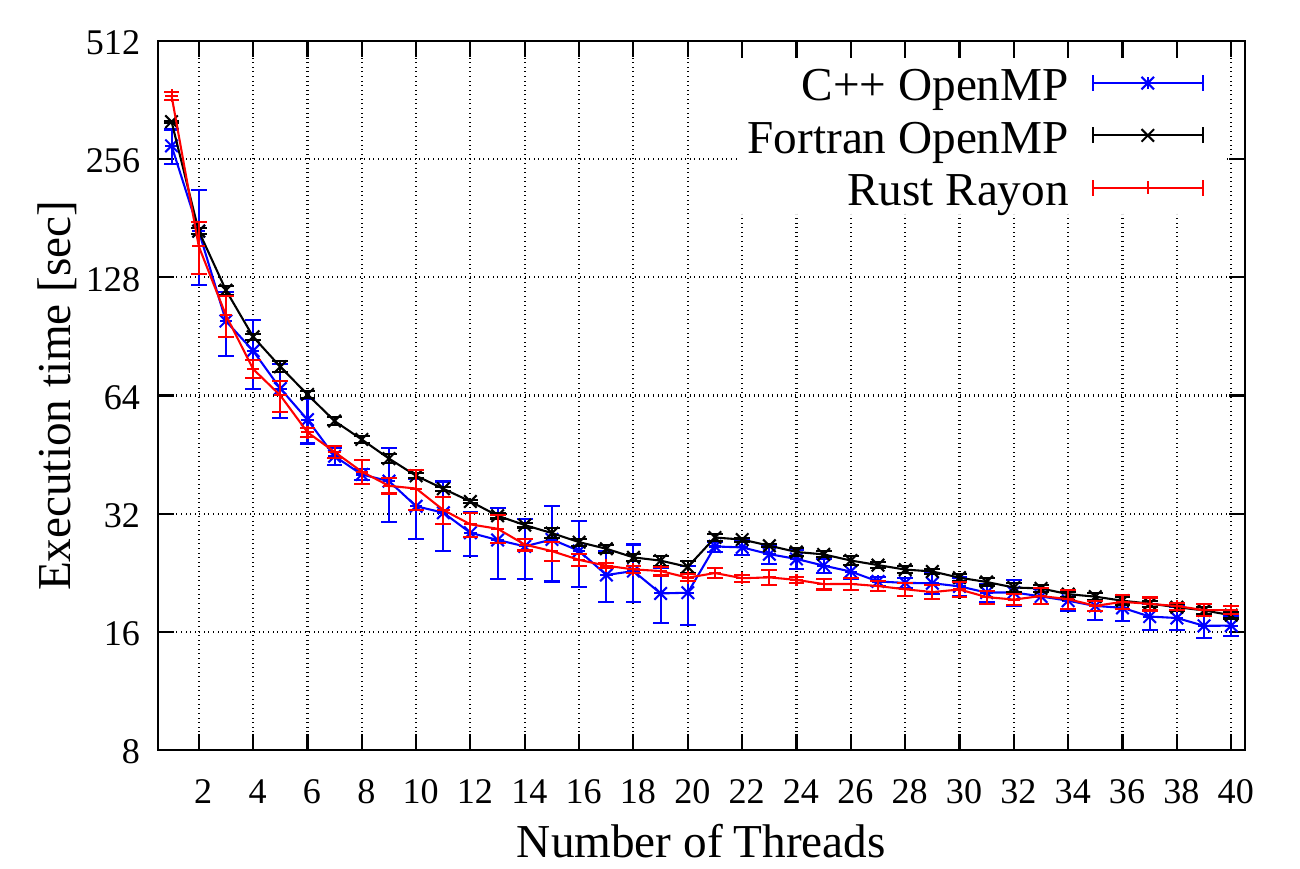}\label{fig:fortranvscpp_ft_b}}\\
         \subfloat[IS]{\includegraphics[width=0.33\textwidth]{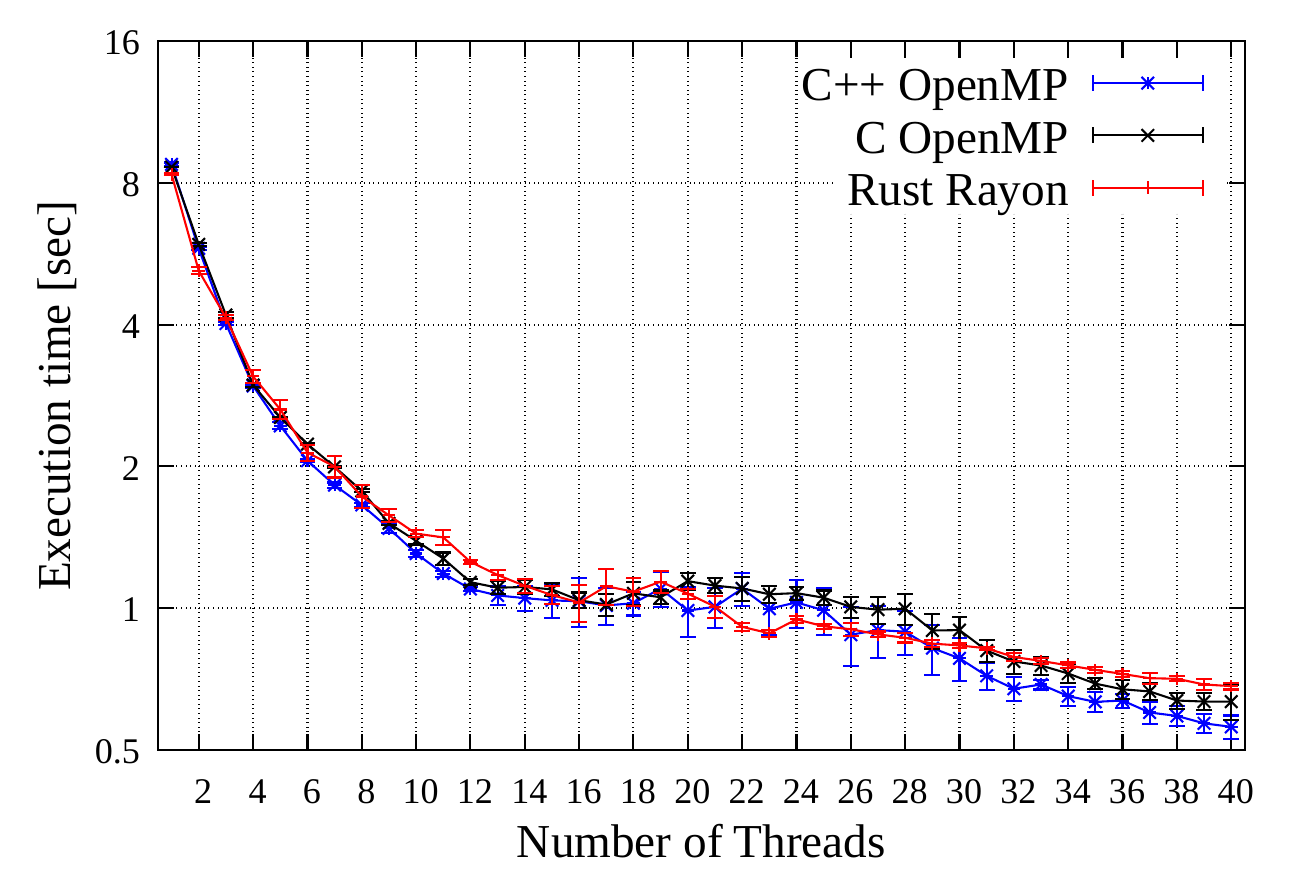}\label{fig:fortranvscpp_is_b}}
         \subfloat[MG]{\includegraphics[width=0.33\textwidth]{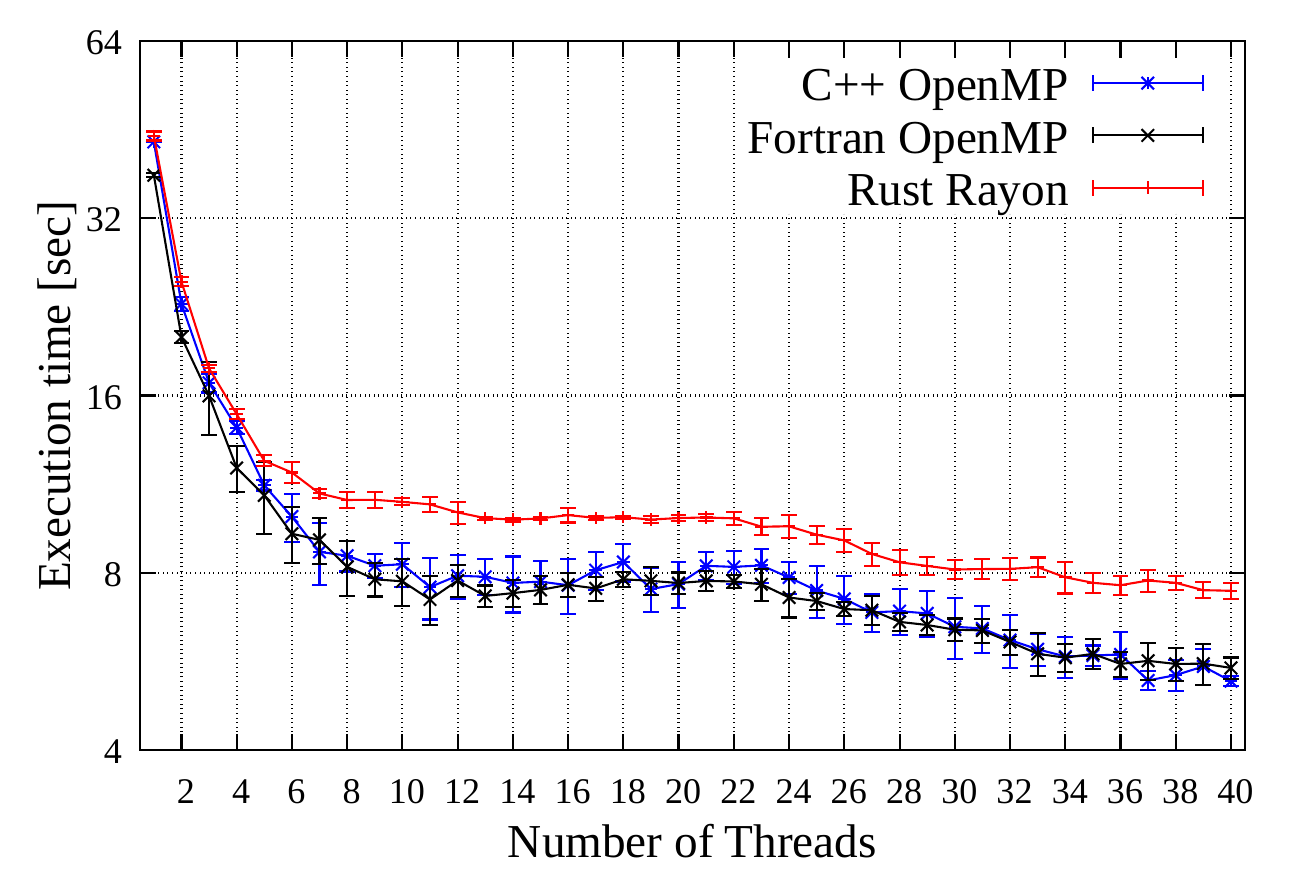}\label{fig:fortranvscpp_mg_b}}
         \subfloat[BT]{\includegraphics[width=0.33\textwidth]{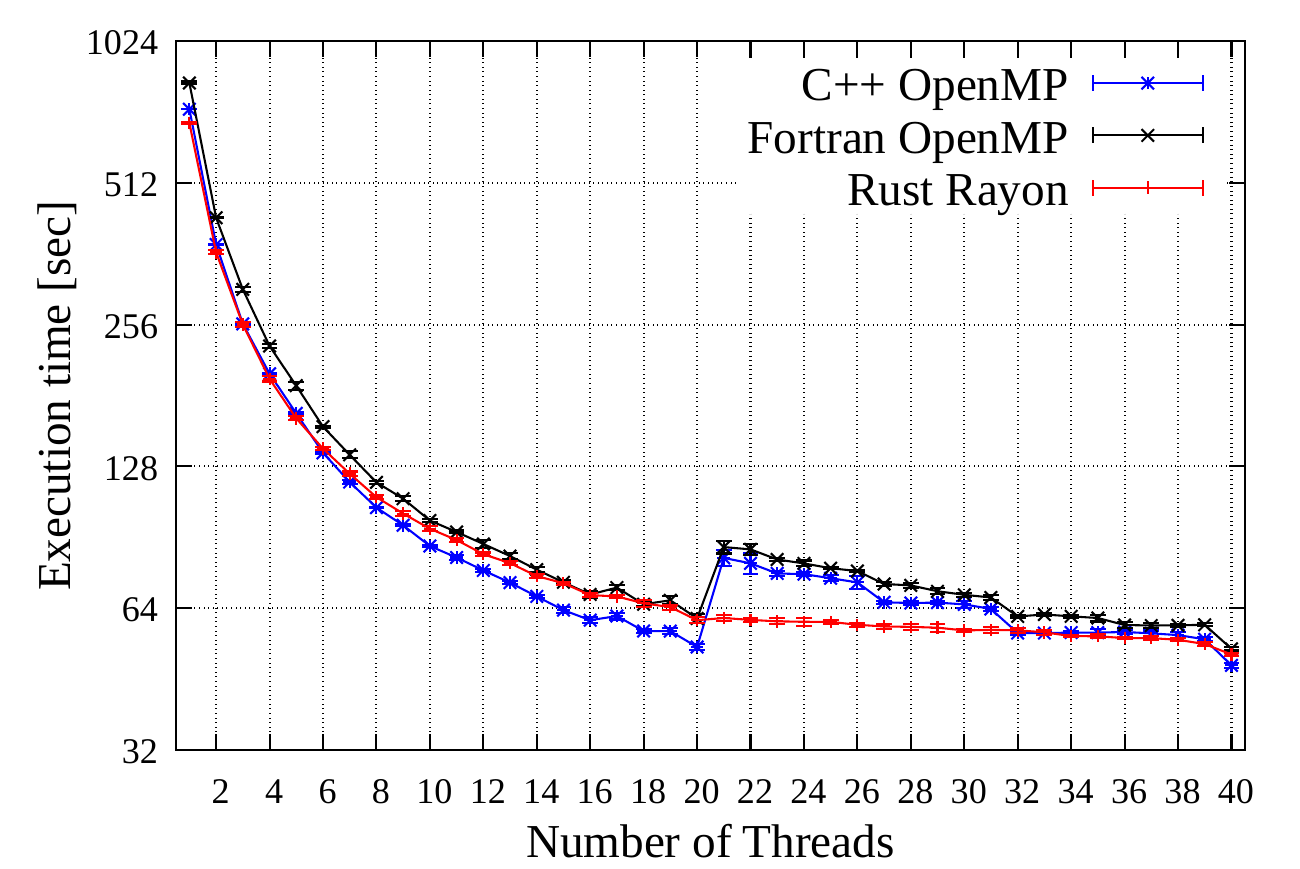}\label{fig:fortranvscpp_bt_b}}\\
         \subfloat[SP]{\includegraphics[width=0.33\textwidth]{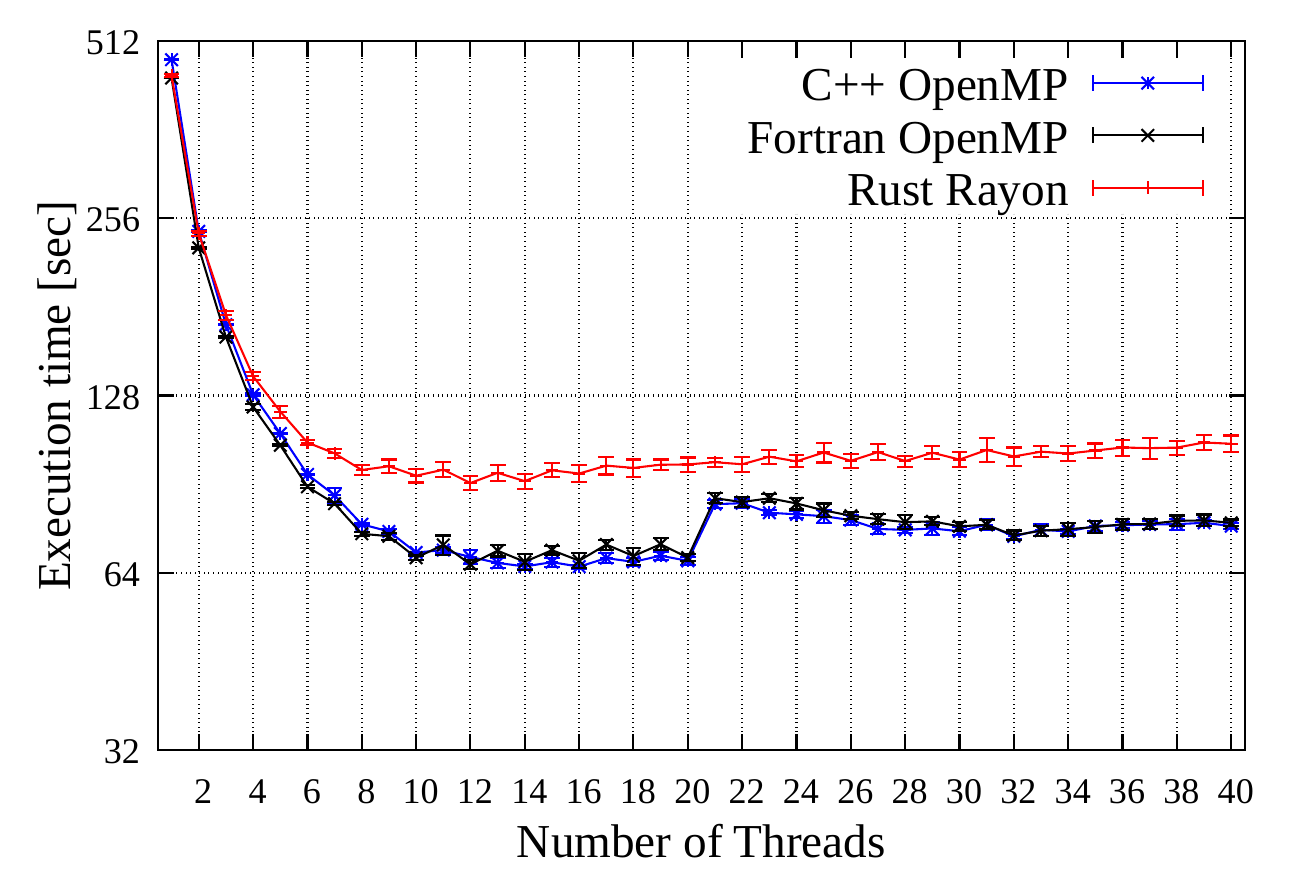}\label{fig:fortranvscpp_sp_b}}
         \subfloat[LU]{\includegraphics[width=0.33\textwidth]{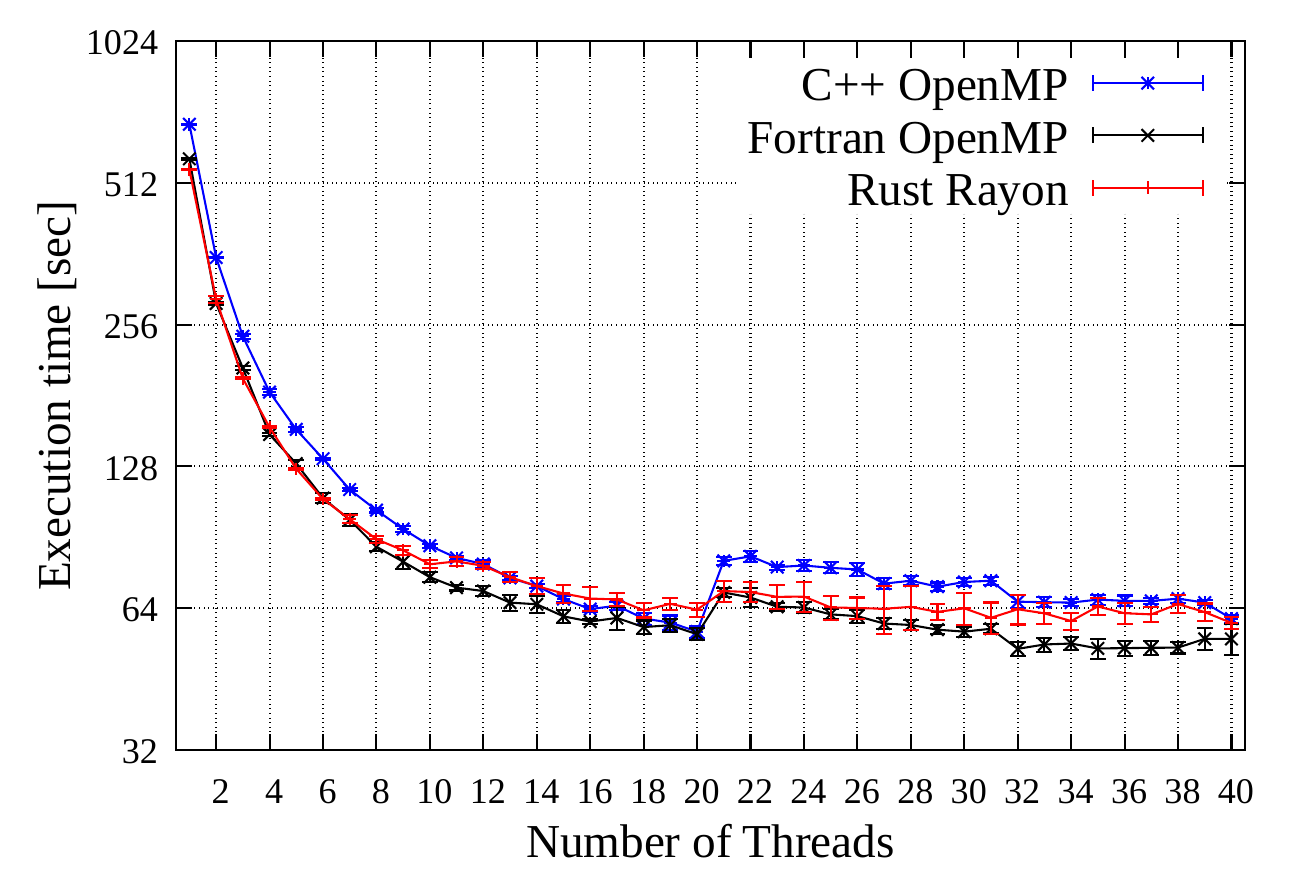}\label{fig:fortranvscpp_lu_b}}
         \caption{Execution time of NPB with OpenMP, NPB-CPP with OpenMP, and NPB-Rust with Rayon for workload class C.}
         \label{fig:graficoslegais}
        \end{figure*}

        \begin{table}[!ht]
        \fontsize{7}{7}\selectfont
        \centering
        \caption{Experimental results showing the best execution time and speedup for each parallel version.}
        \begin{tabular}{C{1.4cm}C{1.4cm}C{1.2cm}C{1.2cm}C{1.2cm}} 
            \toprule
            Benchmark & Metrics & C++ OpenMP & Fortran OpenMP & Rust Rayon \\ 
            \midrule
            EP & 
            \makecell{N Threads\\Time (s)\\Speedup\\Std. Dev.} & 
            \makecell{40\\11.058\\27.677\\0.073} & 
            \makecell{40\\11.132\\28.172\\0.06} & 
            \cellcolor{pastelgreen}\makecell{40\\10.913\\29.666\\0.028}  \\
            \midrule
            CG & 
            \makecell{N Threads\\Time (s)\\Speedup\\Std. Dev.} & 
            \cellcolor{pastelgreen}\makecell{40\\13.682\\22.504\\0.27} & 
            \makecell{40\\12.904\\17.224\\0.408} & 
            \makecell{40\\20.646\\10.45\\0.436}   \\
            \midrule
            FT & 
            \makecell{N Threads\\Time (s)\\Speedup\\Std. Dev.} & 
            \makecell{39\\16.585\\20.163\\1.134} & 
            \makecell{40\\17.646\\18.093\\0.335} & 
            \cellcolor{pastelgreen}\makecell{40\\18.18\\22.152\\0.414}  \\
            \midrule
            IS & 
            \makecell{N Threads\\Time (s)\\Speedup\\Std. Dev.} & 
            \cellcolor{pastelgreen}\makecell{40\\0.56\\14.944\\0.031} & 
            \makecell{39\\0.634\\13.481\\0.025} & 
            \makecell{40\\0.683\\12.308\\0.011}  \\
            \midrule
            MG & 
            \makecell{N Threads\\Time (s)\\Speedup\\Std. Dev.} & 
            \cellcolor{pastelgreen}\makecell{40\\5.239\\8.213\\0.097} & 
            \makecell{40\\5.517\\6.785\\0.229} & 
            \makecell{40\\7.459\\5.103\\0.234}   \\
            \midrule
            BT & 
            \makecell{N Threads\\Time (s)\\Speedup\\Std. Dev.} & 
            \makecell{40\\48.424\\15.367\\0.564} & 
            \cellcolor{pastelgreen}\makecell{40\\52.483\\15.672\\0.42} & 
            \makecell{40\\51.076\\14.925\\0.383}  \\
            \midrule
            SP & 
            \makecell{N Threads\\Time (s)\\Speedup\\Std. Dev.} & 
            \cellcolor{pastelgreen}\makecell{16\\65.573\\7.039\\0.499} & 
            \makecell{12\\66.182\\6.651\\1.137} & 
            \makecell{12\\90.916\\4.878\\2.402}  \\
            \midrule
            LU & 
            \makecell{N Threads\\Time (s)\\Speedup\\Std. Dev.} & 
            \cellcolor{pastelgreen}\makecell{20\\57.116\\11.631\\1.683} & 
            \makecell{32\\52.495\\10.939\\1.878} & 
            \makecell{40\\59.581\\9.007\\1.746}   \\
            \bottomrule
        \end{tabular}
        \vspace{-0.3cm}
        \label{tab:speed}
        \end{table}

    \subsection{Parallel Version}\label{sec:par}

    Figure \ref{fig:graficoslegais} shows the execution time in seconds on a logarithmic scale for each benchmark, varying the number of threads from 1 to 40. Since both NPB and NPB-CPP utilize the OpenMP framework, they exhibit similar results regarding time and scalability. These outputs align with previous tests \cite{NPB-CPP-2021}. In contrast, we observed distinct result patterns in Rust, since Rayon's implementation differs from that of OpenMP. We used the same methodology described in Section \ref{sec:seq} to statistically test the equivalence of the NPB-Rust with Rayon to Fortran and C++ versions with OpenMP. We applied it for each benchmark in each degree of parallelism. The statistical test results indicate that the Rust parallel version is equivalent to Fortran and C++ in 12.5\% and 17.81\% of the cases, respectively.
    
    Rayon's work-stealing and dynamic system scales considerably better compared to OpenMP when Hyper-Threading is enabled. On the other hand, for applications like MG and SP, which are known to stop scaling with fewer threads, Rayon scales similarly to OpenMP at first but hits a limit earlier. The reason stems from the combination of dynamic scheduling and small computation granularity. For the CG benchmark, \texttt{nowait} directives were implemented in the most computationally intensive function in the OpenMP version. Since Rayon does not offer a similar approach for those notations, OpenMP scales better overall in this benchmark. 

    \begin{figure*}[!ht]
         \centering
         \subfloat[1 thread]{\includegraphics[width=0.33\textwidth]{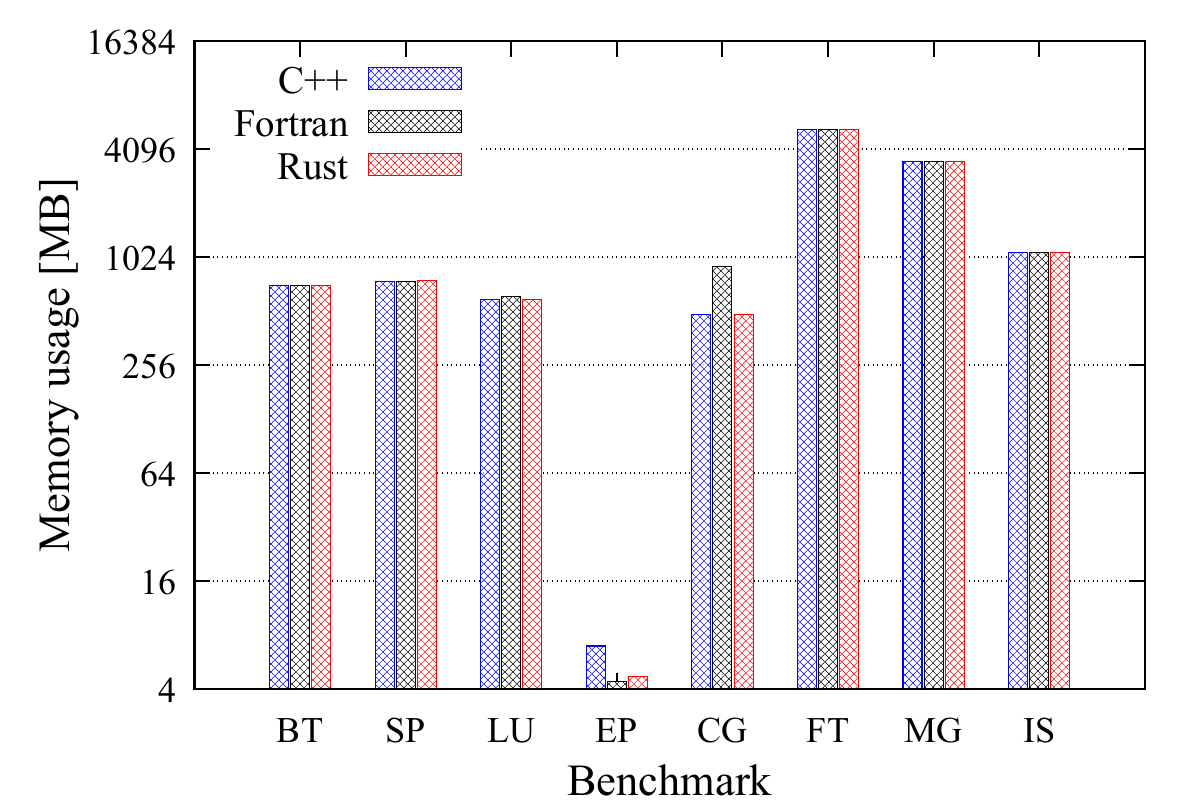}\label{fig:mem1}}
         \subfloat[20 threads]{\includegraphics[width=0.33\textwidth]{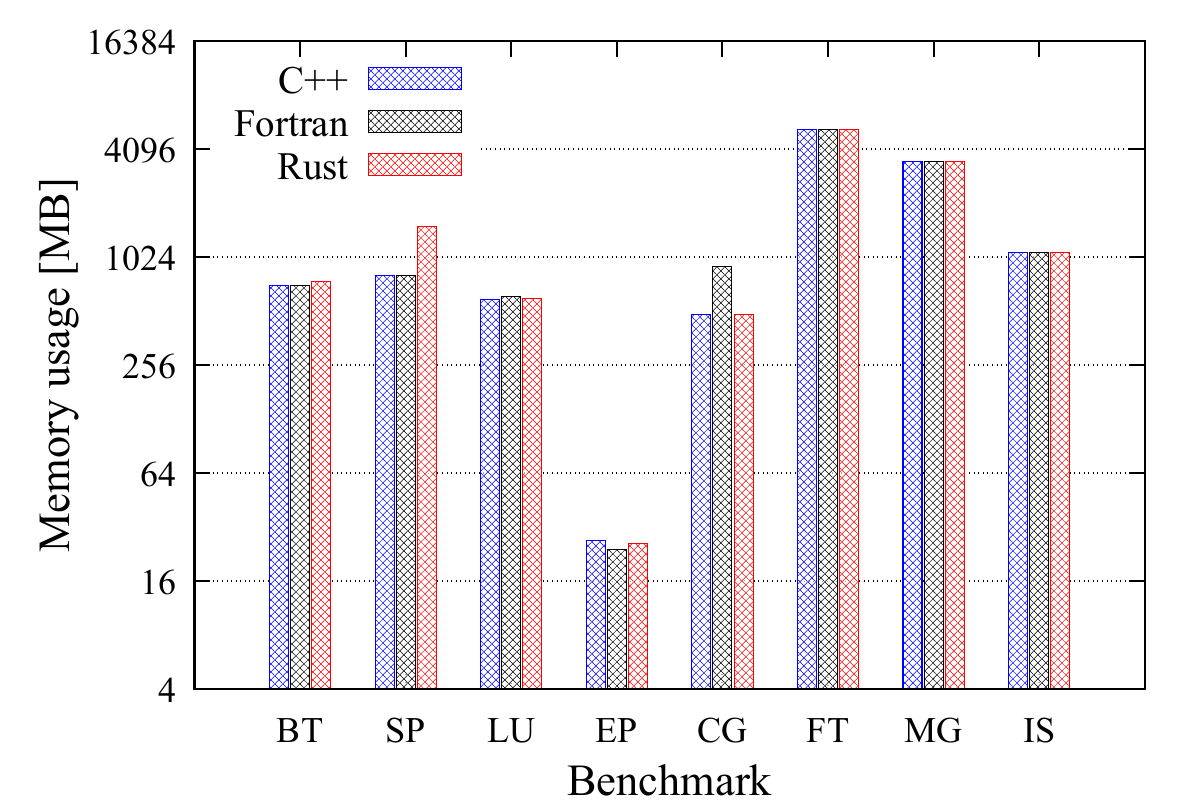}\label{fig:mem20}}
         \subfloat[40 threads]{\includegraphics[width=0.33\textwidth]{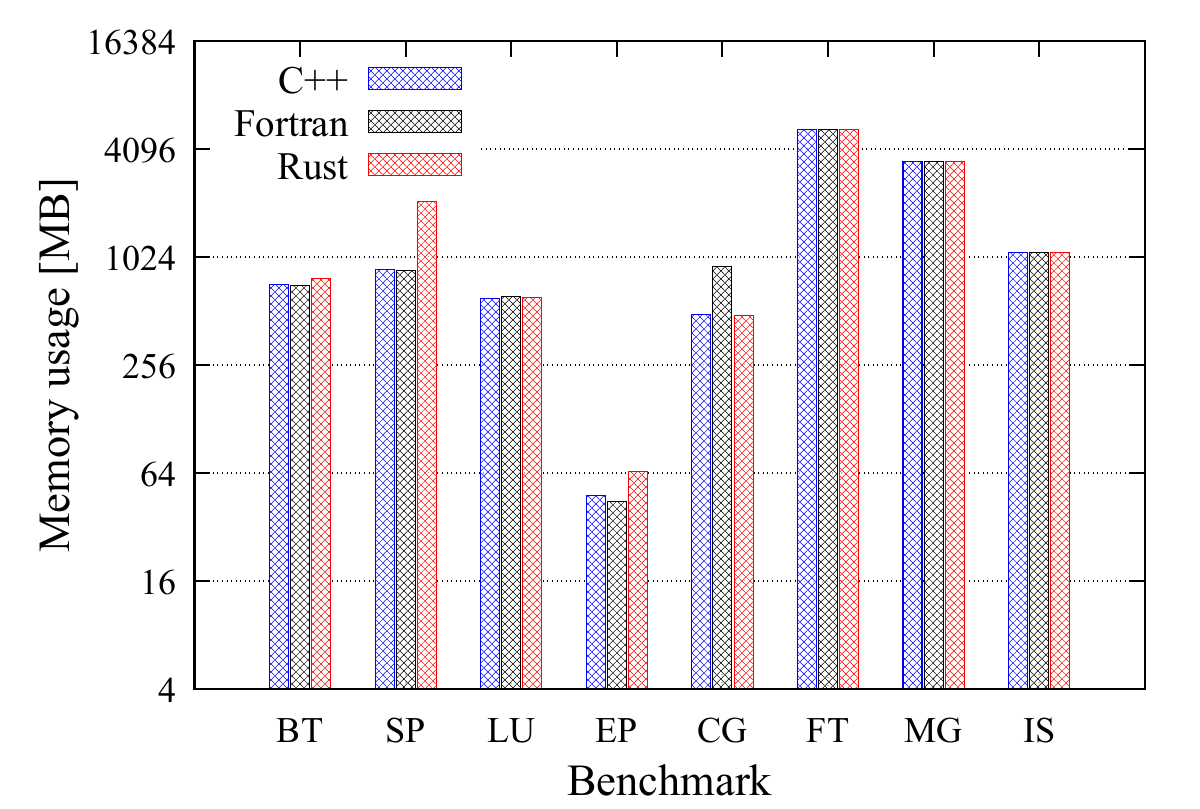}\label{fig:mem40}}
         \vspace{-0.1cm}
         \caption{Memory consumption in MB executing with 1, 20, and 40 threads.}
         \label{fig:mem}
         \vspace{-0.0cm}
        \end{figure*}
    
    Regarding the other applications, Rayon delivered competitive results. Considering the executions using 40 threads for EP, FT, IS, BT, and LU, the geometric mean indicates that Rust with Rayon was 2.74\% slower than the Fortran version and 7.7\% slower than the C++ version with OpenMP. Considering all benchmarks, the difference increases to 16.42\% slower than Fortran and 19.05\% slower than C++. Thus, our hypothesis that Rayon enables the expression of parallel algorithms for NPB with equal performance to OpenMP was refuted.


    We extend the analyses with Table \ref{tab:speed}. It summarizes, for each application and version, the level of parallelism that achieves the best speedup, with green marks highlighting the top one for each benchmark. Rayon managed to reach the highest speedup in the embarrassingly parallel problem due to its work-stealing scheduling. FT is the most memory-intensive benchmark in the NPB suite, which may hide other sources of overhead depending on the platform. Rayon benefits from its dynamic scheduling and reaches the highest speedup on this kernel. As IS execution time is relatively smaller, any runtime overhead has more impact on the results. Although the execution time differences were in order of milliseconds, C++ achieves the best speedup in this kernel. BT was the only application where Fortran with OpenMP achieved the best speedup. All NPB versions presented similar results in this metric for this benchmark. The LU data dependencies in the OpenMP version are handled within the loop scope using flags and \texttt{flush} directives. In contrast, the Rayon version resolves them at the thread scope using locks and conditional variables. These divergent approaches explain the speedup differences.




    Figure \ref{fig:mem} shows the peak of memory consumption in megabytes for each benchmark when using 1, 20, and 40 threads. We observe that most applications maintain similar consumption levels when more threads are enabled. This behavior is expected since the NPB suite shares large data structures between threads in almost all cases. Due to its implementation, the CG kernel in Fortran shows consistently higher memory consumption. The EP kernel demonstrated a relatively higher increase in memory consumption across all versions. Although this kernel naturally has lower memory usage, its parallel version includes a private array to store the random numbers generated for each thread. The SP benchmark in Rust shows considerably higher memory consumption in its highly parallel version. This is related to the increased thread pool stack size configured for the Rust implementation. The OpenMP manages the stack allocation in the solver functions efficiently, while the Rayon version requires more memory.

       \begin{figure*}[t]
         \centering
         \subfloat[Relative SLOC from sequential to parallel versions]{\includegraphics[width=0.5\textwidth]{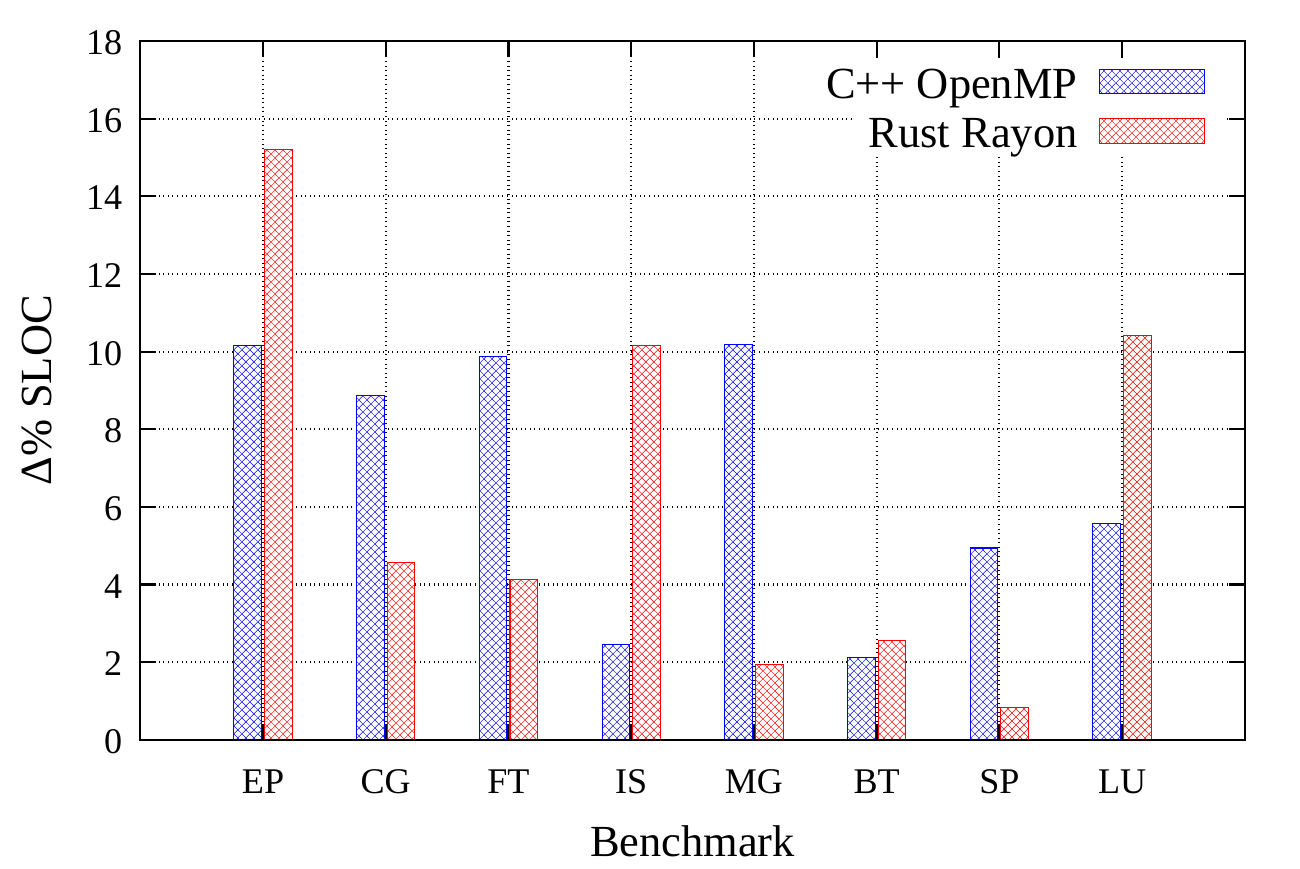}\label{fig:SLOC}}
         \subfloat[Relative COCOMO schedule effort from sequential to parallel versions]{\includegraphics[width=0.5\textwidth]{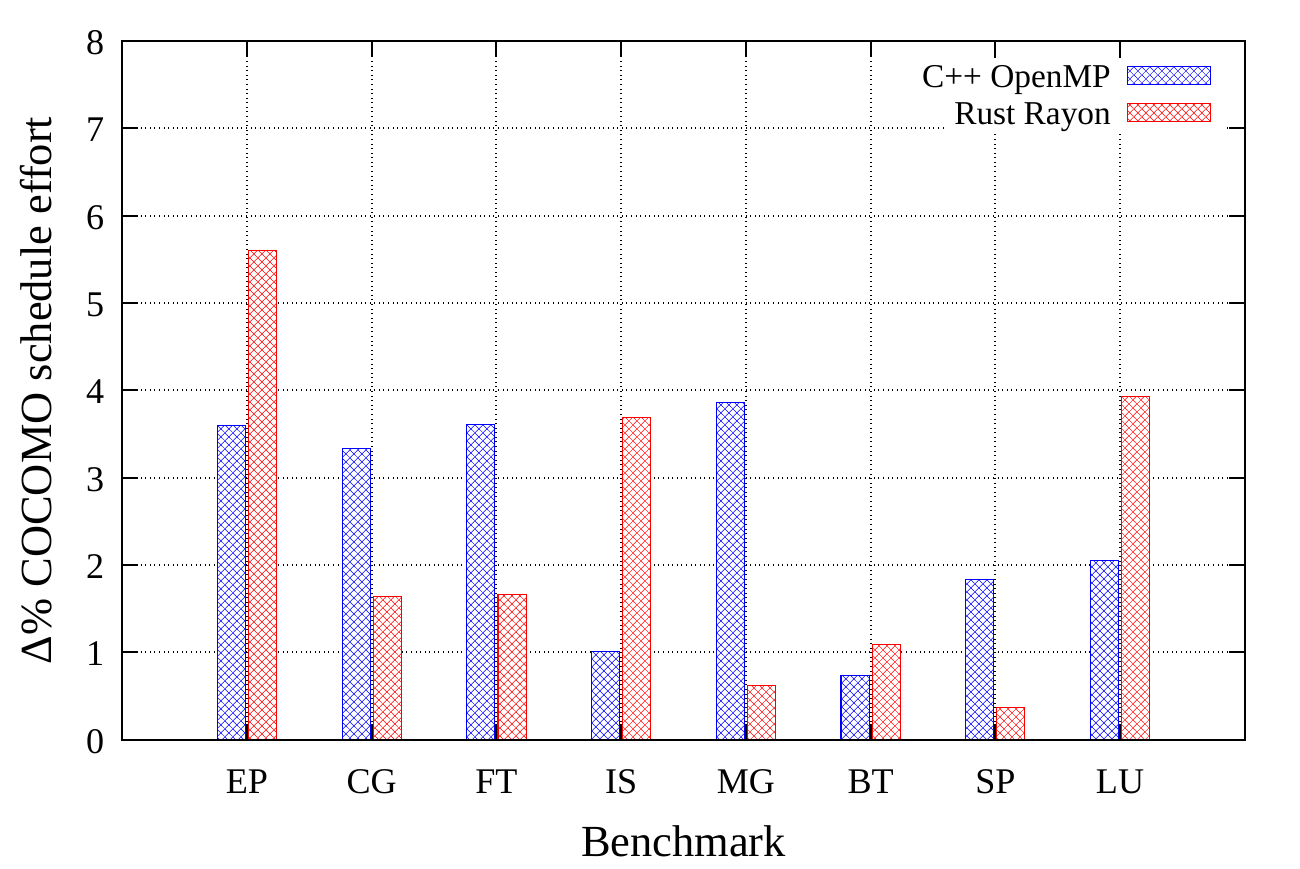}\label{fig:COCO}}
         \caption{Relative SLOC and COCOMO schedule effort from sequential to parallel versions.}
         \label{fig:sloocomo}
         \vspace{-0.3cm}
        \end{figure*}

    To assess the programmability of each framework, significant source lines of code (SLOC) and scheduling efforts based on the constructive cost model (COCOMO) are considered. The scheduling effort is determined based on the lines of code (LOC) and predefined constants customized according to the developer's experience level. In this assessment, we utilized a configuration suitable for developers with high experience. These metrics have already been studied for evaluating parallel programming applications and APIs \cite{program}, and we leverage them for analyzing both parallel and sequential implementations between the C++ and Rust versions. The Fortran version is excluded from this evaluation, as its sequential execution utilizes the same code as the parallel version, but without the OpenMP flag. Figure \ref{fig:SLOC} illustrates the relative SLOC differences between serial and parallel versions of each benchmark, while Figure \ref{fig:COCO} shows the relative differences in scheduling effort. Since COCOMO considers LOC in its calculation, the results for both metrics demonstrate similar patterns. In Rust, EP required a relatively higher effort due to the manual implementation of the reduction function. In the case of LU, the different implementation strategies used to provide organized access to data result in more complexity for Rust with Rayon. IS has relatively more code because of the strategy of manual control over the array access, made to avoid excessive unsafe code. Meanwhile, the other applications needed minimal modifications, as Rayon's iterator-based parallel features align closely with the sequential pattern. The general overview indicates that in the NPB suite, Rayon adds on average 4.6\% more significant source lines of code, while the OpenMP versions in C++ add 5.39\%, respectively. In terms of scheduling effort, Rayon incurs an overhead of 1.75\%, whereas OpenMP introduces an overhead of 2.09\% in C++. From the NPB perspective, Rayon and OpenMP frameworks have a minimal impact on the programmability efforts.






\section{Related Work}\label{sec:rw}
\textbf{Rust performance.} Over the past decade, studies have investigated the applicability of Rust for HPC, aiming to achieve safety without compromising performance. Rust with CUDA for general-purpose GPU programming was compared to C++ in matrix multiplication and array copying \cite{rw_tes}. The findings indicated that Rust provides competitive performance compared to C++, with minimal reliance on unsafe code. The performance of Rust's data structures, including HashMap and binary tree, and simple algorithms like merge sort and insertion sort, was also compared to C++ \cite{rw_isfast}. Results demonstrate that dictionary operations and insertion sort are faster in C++, while Rust outperforms C++ in merge sort. The N-body problem was implemented in Rust with Rayon to compare its MFLOPS performance against C with OpenMP \cite{rw_pervsprog}. The results show that C outperformed Rust when using single-precision operations due to more efficient mathematical optimizations during assembly generation. However, when using double precision, Rust's performance was similar to C's. Considering the simplicity of the benchmarks tested in these works within the HPC context, further analysis could be extended with more complex applications, such as the NPB.

\textbf{Benchmarks.} The applications from the Problem-Based Benchmark Suite (PBBS) \cite{pbbs_c} were ported from C++ to Rust \cite{rw_zerocost}. The study focused on examining Rust's capabilities in handling regular and irregular parallelism and evaluating the impact of using unsafe. Rust performed slightly better in the sequential version, while C++ with OpenMP outperformed Rust with Rayon. They highlight Rust's ability to handle read-only and regular parallelism, but when dealing with irregular parallelism, unsafe was often necessary to avoid runtime overheads. Still, the NPB is distinguished by its focus on intensive arithmetic tasks and complex scientific applications derived from CFD. The researchers in \cite{PIEPER:COLA:21} created a benchmark to perform a set of experiments in stream processing and data-parallel Rust applications. They compared different Rust parallel programming interfaces and assessed their parallelism performance. In contrast, we focus on more robust NPB applications and compare them with Fortran and C++.

\textbf{NPB in prior works.} The NPB has been widely used for benchmarking different programming languages, frameworks, and parallel programming strategies. In this work, we leverage NPB-CPP \cite{NPB-CPP-2021}, though many other variants exist, such as NPB-MPJ \cite{NPB-MPJ}, which utilizes Java's message-passing interfaces. The NPB suite is frequently employed to evaluate GPU frameworks for parallelism. A Python version was implemented based on the NPB-CPP to test the Numba environment for CUDA \cite{NPB-PY}. Other implementations have been proposed for OpenCL \cite{npb-opencl} and evaluated against the original OpenMP parallel version. The NPB also has independent versions for OpenACC \cite{npb-openacc} and CUDA \cite{NPB-CUDA}. New implementations of varying parallelism strategies and frameworks often emerge. Therefore, NPB-Rust holds significant value for the research community, enabling Rust to be included in future evaluations.








\section{Conclusion}\label{sec:conl}

This paper contributed with NPB-Rust, a Rust-based benchmark suite designed to represent intensive arithmetic tasks and complex scientific applications in Rust. Along with a sequential version, we developed a parallel implementation using the Rayon library. We conducted a performance evaluation focusing on execution time, scalability, speedup, memory consumption, and programmability. This analysis compared the sequential and parallel implementations of NPB-Rust with the NPB versions in Fortran and C++, both parallelized with the OpenMP framework. Rust has been demonstrated to be suitable for expressing NPB applications, but it required unsafe Rust to achieve better performance on irregular parallelism scenarios and for other specific optimizations. When testing our hypothesis, we found that Rust outperformed C++ in the sequential code but was slower than Fortran. Therefore, we tested the NPB-Rust, and it showed reliable performance. Although Rayon presented a competitive performance with OpenMP in most of the cases and also provides the benefit of memory safety, experimental results and hypothesis test indicates that Rayon was slower overall in the NPB suite. Thus, there are opportunities for exploring different parallelism methodologies to achieve better parallelism exploitation with Rust in the future. We plan to continually enhance the NPB-Rust suite by implementing it with other Rust parallel programming interfaces and experimenting with other parallel architectures over different workloads.



\section*{Acknowledgment}

This research was supported by Coordena\c{c}\~{a}o de Aperfei\c{c}oamento de Pessoal de N\'{i}vel Superior - Brasil (CAPES) - Finance Code 001, FAPERGS 09/2023 PqG (N\textsuperscript{o} 24/2551-0001400-4), and CNPq Research Program (N\textsuperscript{o}306511/2021-5).




\bibliographystyle{ieeetr}
\bibliography{Bib/bib}


\end{document}